\documentclass[aps,pra,twocolumn,superscriptaddress,showpacs,amssymb,amsmath]{revtex4}
\usepackage{graphicx}
\usepackage{amsmath}
\begin{document}

\title{Localization of wave packets in one-dimensional random potentials}
\author{Juan Pablo Ram\'irez Valdes}
\author{Thomas Wellens}
\affiliation{Physikalisches Institut, Albert-Ludwigs Universit\"at Freiburg, Herman-Herder-Str. 3a, 79104 Freiburg, Germany}
\date{\today}

\begin{abstract}
We study the expansion of an initially strongly confined wave packet in a one-dimensional weak random potential with short correlation length. At long times, the expansion of the wave packet comes to a halt due to destructive interferences leading to Anderson localization. We develop an analytical description for the disorder-averaged localized density profile. For this purpose, we employ the diagrammatic method of Berezinskii which we extend to the case of wave packets, present an analytical expression of the Lyapunov exponent which is valid for small as well as for high energies and, finally, develop a self-consistent Born approximation in order to analytically calculate the energy distribution of our wave packet. By comparison with numerical simulations, we show that our theory describes well the complete localized density profile, not only in the tails, but also in the center.
\end{abstract}

\maketitle

\section{Introduction}
Transport phenomena in mesoscopic and disordered systems may be severely affected by interferences due to the wave nature of quantum theory. This is especially the case for one-dimensional systems, where even a weak random potential is known to lead to Anderson localization \cite{Anderson}. In the limit of long times, the particle then remains localized in a finite region surrounding its initial position. The main characteristic of Anderson localization is the exponential decay of the probability density of eigenstates in the random potential, where the corresponding Lyapunov exponent defines the inverse localization length and depends on the corresponding energy \cite{Alberto}. The dynamics of a wave packet in a one-dimensional random potential is a more complex phenomenon due to the superposition of many eigenstates with different energies.
%tw which, 
In general, this %tw
leads to a non-exponential decay of the asymptotic density profile reached at long times 
%tw \cite{Sanchez, Errsanchez}. 
\cite{Sanchez}.
How to describe the localized disorder-averaged asymptotic probability density is
%tw
therefore %tw
 a relevant theoretical problem in order to understand the suppression of transport due to the presence of disorder. 

%tw 
In the last decade, several experiments on the expansion of Bose-Einstein condensates  in one-dimensional random potentials have been performed \cite{Lye,Clement,Fort,Clement2,Billy}. 
These experiments have been interpreted in terms of Anderson localization of non-interacting particles, although the interactions between the particles forming the condensate do affect the shape of the initial state reached during a short expansion period following the release of the initially trapped condensate, after which the interactions are assumed to be negligibly small. The appearance of the disorder potential then localizes the condensate's density distribution \cite{Piraud,Sanchez2,Pezze}. Theoretical descriptions have been developed which explain well the exponential or algebraic decay of the localized density profile in the far tails (i.e. for large distances) 
%tw \cite{Sanchez, Errsanchez,Piraud}, 
\cite{Sanchez, Piraud}, 
but, so far,  no theory has been able to reproduce the 
%tw
behaviour at the %tw
center 
%tw of the density profile 
close to the initial position, where the maximum of the density 
%tw
profile %tw
remains. 
 
To fill this gap is the main purpose of the present article. We will present a theoretical description of the complete density profile (including the center and the tails) which agrees well with the results of numerical calculations. 
%tw In order 
To capture the correct behaviour at the center, it is essential to treat, in particular, the regime of small energies in an adequate way. We achieve this aim by
developing a self-consistent Born approximation in order to calculate the spectral function, as well as by making use of an exact analytical expression of the Lyapunov exponent for uncorrelated potentials \cite{Thouless2}, which is valid for large as well as for small energies.

The article is organized as follows: First, we introduce the model used throughout the paper to describe the expansion of a wave packet  in a one-dimensional random potential in Sec.~\ref{sec:model}. Then, our theoretical approach will be presented in Sec.~\ref{sec:theory}.
By applying the diagrammatic theory of Berezinskii \cite{Berezinskii} to the case of wave packets in Sec.~\ref{subsec:heuristic}, we derive a simple equation which expresses 
the 
%tw desired spatial 
localized %tw
density profile as an integral over the density correlation function already known in the literature for the case of fixed energy \cite{Gogolin,Gogolin2}, weighted with the energy distribution of our wave packet.  The density correlation function, in turn, depends on the Lyapunov exponent characterizing the exponential decay of energy eigenfunctions, which we will calculate in Sec.~\ref{sec:lyapunov} for our correlated random potential. In order to find an analytical expression for the energy distribution, we will develop a suitable version of the self-consistent Born approximation in Sec.~\ref{subsec:spectraldensity}, which, unlike the ordinary Born approximation, does not exhibit a divergence at small energies. In Sec.~\ref{sec:results}, our theory will be tested by comparison with numerical calculations. Sec.~\ref{sec:conclusion} concludes the paper. Some technical aspects concerning the extension of Berezinkii's diagrammatic method to the case of wave packets are  
relegated to Appendix~\ref{sec:diagrammaticderivation}.

\section{Model and basic assumptions}
\label{sec:model}

To describe the propagation of non-interacting bosons in a one-dimensional random potential, we use the single-particle Hamiltonian:
\begin{equation}
\label{eq:H}
H=p^2+V(x),
\end{equation}
with momentum operator $p$ and random potential $V(x)$. Note that, throughout this paper, we use units such that $\hbar=2m=1$.

The random potential is modeled as a Gaussian random process, with vanishing mean value $\overline{V(x)}=0$ and two-point correlation function:
\begin{equation}
\label{correlationx}
C_2(x'-x)=\overline{V(x')V(x)}=V^2_0 e^{-|x'-x|^2/(2\sigma_c^2)},
\end{equation}
defining the strength $V_0$ of the random potential and its correlation length $\sigma_c$. The over-line in Eq.~(\ref{correlationx}) represent the average over all realizations of the random potential. In the following, we assume that either the strength or the correlation length (or both) of the random potential is small, such that $V_0\sigma_c^2\ll 1$. This condition implies that the kinetic energy of a particle with wavelength comparable to the correlation length is much larger than the 
%tw
typical size $V_0$ of the potential's %tw 
fluctuations.
%tw of the potential. 
In Sec.~\ref{sec:lyapunov}, this will  allow
%tw This allows 
us either to treat $V(x)$ as a perturbation (in the regime of large energies), or to approximate $V(x)$  by an uncorrelated potential (in the regime of low energies, where, due to the long wavelength, the particle is not able to resolve the spatial correlations).

As initial state, we consider a Gaussian wave packet:
\begin{equation}
\label{initialgaussianstate}
\psi_0(x)=\langle x|\psi_0\rangle=\left(\frac{1}{\pi a^2}\right)^{1/4}e^{-x^2/(2a^2)},
\end{equation}
%tw factor 2! %tw
where $a$ is the initial width of the wave packet, which we assume to be much smaller than the localization length
%tw
(see Sec.~\ref{sec:lyapunov}) %tw
 in the relevant range of energies 
 %tw
 (see Sec.~\ref{subsec:spectraldensity}). %tw
 For the numerical solution of the wave packet propagation induced by Eq.~(\ref{eq:H}), we use periodic boundary conditions, i.e. $\psi(x+L,t)=\psi(x,t)$ with system size $L\gg a,\sigma_c$.

Throughout this paper, we will be interested in the disorder-averaged asymptotic particle density assumed at long times $t$, which is defined by:
\begin{equation}
\label{asymdef}
\overline{n(x)}=\lim_{T \to \infty}\frac{1}{T}\int_{0}^{T} {\rm d}t \ \overline{n(x,t)},
\end{equation}
where:
\begin{equation}
\label{eq:av_density}
\overline{n(x,t)}=\overline{|\langle x|e^{-i \hat{H}t}|\psi_0\rangle|^2}
\end{equation}
denotes the average density at position $x$ and time $t$.
It is useful to write $\overline{n(x)}$ in terms of the eigenfunctions $\{|\phi_n\rangle\}$ of the Hamiltonian $H$:
\begin{equation}
\label{asymdefeigenfunctions}
\overline{n(x)}=\overline{\sum_{n=0}^{\infty}|\langle x|\phi_n\rangle\langle\phi_n|\psi_0\rangle|^2}.
\end{equation}
Eq.~(\ref{asymdefeigenfunctions}) is valid under the assumption of a non-degenerate energy spectrum $\{E_n\}$, which is true for almost all realizations of the random potential, since the probability of encountering a single realization $V(x)$ of the random potential with degenerate spectrum vanishes.
 
\section{Theoretical approach}
\label{sec:theory}

As mentioned in the introduction, the approach that we will follow in order to achieve a theoretical and, as far as possible, analytical description of   the asymptotic density $\overline{n(x)}$ is based on the diagrammatic method of Berezinskii \cite{Berezinskii}. In contrast to our case, Berezinskii considers the case of electronic (i.e. fermionic) transport which is mediated by only those eigenstates the energy of which is close to the Fermi energy $E_F$. In addition, he does not use a wave packet as initial state, but rather calculates density-density correlations between two definite  points $x'$ and $x$, respectively. As compared to Eq.~(\ref{asymdefeigenfunctions}), the quantity calculated by Berezinskii can be defined as:
\begin{equation}
\label{asymdefeigenfunctionsberezinskii}
\overline{n_E(x-x')}=\overline{\sum_{n=0}^{\infty}|\langle x|\phi_n\rangle\langle\phi_n|x'\rangle|^2\delta(E_n-E)}/\rho_E,
\end{equation}
(with $E=E_F$) where the density of states: 
\begin{equation}
\label{eq:rho}
\rho_E=\overline{\sum_n |\langle\phi_n|x'\rangle|^2\delta(E_n-E)}, 
\end{equation}
is 
%tw required for
introduced to ensure
 normalization, i.e. $\int {\rm d}x~\overline{n_E(x)}=1$. As shown in \cite{Berezinskii}, this quantity, which we will call \lq asymptotic state at fixed energy\rq\ in the following, is related to the density-density correlation function of non-interacting electrons at zero  temperature.

In order to generalize Berezinskii's approach to the case of wave packets, we will first derive the following relation between our $\overline{n(x)}$ and
Berezinskii's $\overline{n_E(x-x')}$: 
\begin{eqnarray}
\overline{n(x)} & = & \int_{-\infty}^\infty {\rm d}E\int_{-\infty}^\infty {\rm d}p {\rm d}q\ W(q,p) A(p,E)\nonumber\\
& & \times \sum_{j=-\infty}^\infty \overline{n_{E}(x-q+j L)}, 
\label{density}
\end{eqnarray}
which takes into account the position and momentum uncertainty of the initial state  as given by its Wigner function $W(q,p)$, see Eq.~(\ref{eq:wigner}), as well as the resulting energy distribution due to the spectral function $A(p,E)$, see Sec.~\ref{subsec:spectraldensity}. Finally, the periodic boundary conditions (with period $L$) are accounted for by the sum over $j$ in the last line.

For the case of an initially strongly confined wave packet, where the initial width $a$ is much smaller than the final width of the asymptotic state $\overline{n(x)}$, we simplify Eq.~(\ref{density}) by neglecting the dependence of $\overline{n_{E}(x-q+j L)}$ on the initial position $q\simeq 0$:
\begin{equation}
\label{densitysimple}
\overline{n(x)} = \int_{-\infty}^\infty {\rm d}E~P(E)~\sum_{j=-\infty}^\infty \overline{n_{E}(x+j L)},
\end{equation}
where:
\begin{equation}
\label{PE}
P(E)=\int_{-\infty}^\infty {\rm d}q{\rm d}p \ W(q,p) A(p,E)
\end{equation}
denotes the energy distribution of our wave packet.

Although Eq.~(\ref{density}) appears to be plausible from a physical point of view -- and has 
%tw
already %tw
been used,
%tw
e.g., %tw
 in \cite{Piraud} as a natural ansatz without further explanation -- a rigorous justification of
Eq.~(\ref{density}) has so far been lacking. In the following Sec.~\ref{subsec:heuristic} (together with Appendix~\ref{sec:diagrammaticderivation}), we will therefore present a diagrammatic derivation of Eq.~(\ref{density}) following the method of  Berezinskii \cite{Berezinskii}. The subsequent Secs.~\ref{sec:lyapunov} and \ref{subsec:spectraldensity} will then explain the theoretical approaches which we have developed in order to determine the fundamental quantities entering our description of the asymptotic density according to Eq.~(\ref{density}), i.e. the spectral function $A(p,E)$ and the asymptotic state at fixed energy $\overline{n_E(x)}$.

\subsection{Diagrammatic description of the asymptotic average density}
\label{subsec:heuristic}

We first give a short introduction into the diagrammatic method of Berezinskii, originally developed in \cite{Berezinskii} for white noise potentials, and later extended by Gogolin \cite{Gogolin,Gogolin2} to the case of correlated random potentials with short correlation length. As in these original works, we consider an infinitely extended system (without periodic boundary conditions) in this subsection.

\subsubsection{The diagrammatic method of Berezinskii}
\label{subsec:berezinskii}

\begin{figure}[h]
\includegraphics[width=8.5cm]{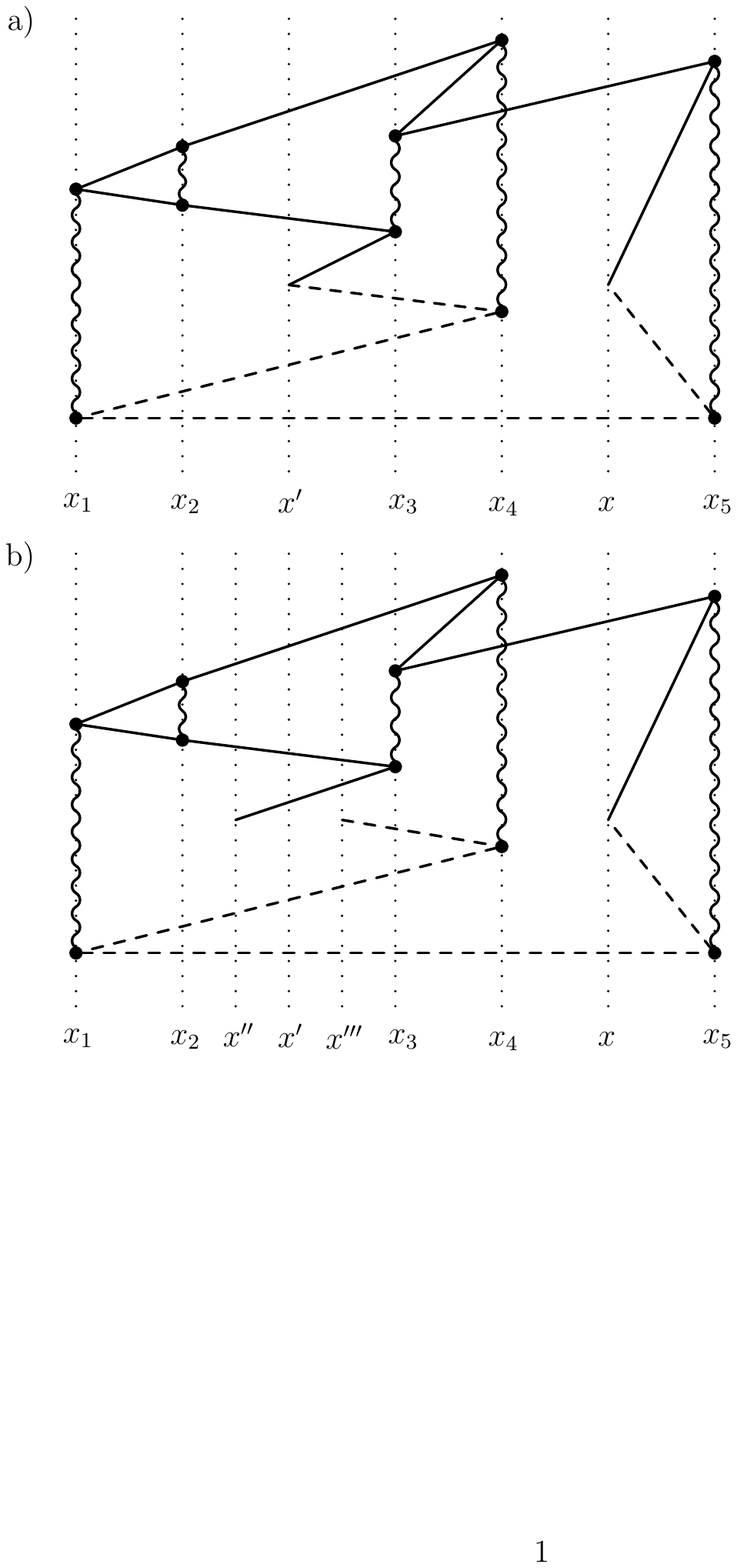}
\caption{\label{fig:berezinskii} a) Example of a diagram which contributes to the average product  $\overline{G^{(+)}(x,x',E+\omega)G^{(-)}(x',x,E)}$ of Green functions in Berezinskii's approach. The solid (or dashed) lines represent
free-particle Green functions $G_0^{(+)}(E+\omega)$ (or $G_0^{(-)}(E)$), respectively. The dots connected by vertical wavy lines denote the
two-point correlation function of the random potential.
b) Diagram contributing to $\overline{G^{(+)}(x,x'',E+\omega)G^{(-)}(x''',x,E)}$ for the case of wave packets. The diagram is almost identical to the one shown in a), except for the fact that
the source point $x'$ of a) is replaced by two different source points $x''$ and $x'''$. As compared to a), this introduces an additional  factor $e^{i p_E(x'''-x'')}$.
}
\end{figure}

The diagrammatic method relies on the fact that the asymptotic density at fixed energy, see Eq.~(\ref{asymdefeigenfunctionsberezinskii})
%tw
(but without the average), can be expressed as follows:
\begin{eqnarray}
n_E(x-x') & = & \lim_{\omega \to 0} \frac{\omega}{2\pi i\rho_E} G^{(+)}(x-x',E+\omega)\nonumber\\
& & \hspace{-1cm}\times\left(G^{(-)}(x'-x,E)-G^{(+)}(x'-x,E)\right),
\label{bereft0}
\end{eqnarray}
%tw new equation!
in terms of retarded and advanced Green functions:
\begin{equation}
\label{greenbere}
G^{(\pm)}(x-x',E)=\langle x|[E-H\pm i\eta]^{-1}|x'\rangle,
\end{equation} 
where $\eta$ is an infinitesimal, positive quantity. Due to translational invariance, $G^{(\pm)}(x-x',E)$ depends only on the difference $x-x'$ between the final and the initial point $x$ and $x'$, respectively.
%tw
Taking the disorder average of Eq.~(\ref{bereft0}), the product of two retarded Green functions
can be assumed to vanish due to the random phase of $G^{(+)}(x-x',E)$. In contrast, the product of a retarded and an advanced Green function survives the disorder average since the phases of $G^{(+)}(x-x',E+\omega)$ and $G^{(-)}(x'-x,E)$ compensate each other (in the limit $\omega\to 0$) in each single realization of the random potential. This leads to:
\begin{equation}
\label{bereft}
\overline{n_{E}(x-x')}=  \lim_{\omega \to 0} \frac{\omega}{2\pi i\rho_E} \overline{G^{(+)}(x-x',E+\omega)G^{(-)}(x'-x,E)}.
\end{equation}
%tw Eq.~(\ref{bereft}) can be derived from Eq.~(\ref{asymdefeigenfunctionsberezinskii}) by neglecting the product $\overline{G^{(+)}(x-x',E+\omega)G^{(+)}(x'-x,E)}$ of two retarded Green functions, which is assumed to vanish on average 
Then, the Green functions are expanded in powers of the random potential $V$:
\begin{equation}
G^{(\pm)}(E)=G_{0}^{(\pm)}(E)+G_{0}^{(\pm)}(E) V G_{0}^{(\pm)}(E)+\dots,
\end{equation}
with  free-particle Green functions $G_{0}^{(\pm)}(E)$. Thus, the Green functions $G^{(+)}(E)$ and $G^{(-)}(E)$ are written as a sum of infinitely many terms, each of which can be represented as a diagram, where retarded (advanced) Green functions are represented, e.g., by solid (dashed) lines, and the random potential $V$ by a dot. Performing the disorder average of the product $G^{(+)}(E+\omega)G^{(-)}(E)$, dots become connected by wavy lines indicating two-point correlation functions of the random potential (also called \lq vertices\rq\ in the following). An example of such a diagram contributing to the average product $\overline{G^{(+)}(E+\omega)G^{(-)}(E)}$ is given in Fig.~\ref{fig:berezinskii}.
The corresponding mathematical expression is given in Eq.~(\ref{eq:example_berezinskii}) in Appendix~\ref{sec:diagrammaticderivation}.

The next important approximation consists of assuming that -- similar to the neglect of $\overline{G^{(+)}(E+\omega)G^{(+)}(E)}$ discussed above --
only those diagrams survive the disorder average where the phase factors associated with the free-particle Green functions exactly compensate each other in the limit $\omega\to 0$. This is the case if each space interval between neighbouring vertices contains the same number of retarded and advanced Green functions. In Fig.~\ref{fig:berezinskii}, for example, there are two solid and two dashed lines between the points $x_2$ and $x'$, whereas there are three solid and three dashed lines between $x'$ and $x_3$, etc. We will call a diagram of this type \lq essential diagram\rq.
As shown in Appendix~\ref{sec:diagrammaticderivation}, they can be systematically constructed in terms of certain elementary vertices displayed in Fig.~\ref{fig:vertices}.

After these introductory steps, the main part of Berezinskii's paper is concerned with performing the sum over all these diagrams, which yields the desired analytical expression for  $\overline{n_{E}(x-x')}$. These steps will be explained in Appendix~\ref{sec:diagrammaticderivation}. Note that a crucial element of Berezinskii's derivation is the ability to introduce a spatial ordering between the positions of vertices (i.e. $x_1<x_2<x'<x_3<x_4<x<x_5$ in Fig.~\ref{fig:berezinskii}). For this reason, the method of Berezinskii is restricted to one-dimensional systems and cannot be generalized to higher dimensions. Berezinskii's result has been simplified by Gogolin \cite{Gogolin,Gogolin2} who finally arrives at:
\begin{eqnarray}
\label{eigenlocstate}
\overline{n_{E}(x-x')}&=&\frac{\pi^2\gamma(E)}{8}\int_0^{\infty} du\ u \sinh{(\pi u)}\left[\frac{1+u^2}{1+\cosh{(\pi u)}}\right]^2 \nonumber\\
&\times&\exp\{-(1+u^2)\gamma(E)|x-x'|/2\}.
\end{eqnarray}
Here, $\gamma(E)$ denotes the energy-dependent Lyapunov exponent (the inverse of the localization length). Berezinskii's approach
reproduces the Born approximation for the Lyapunov exponent, i.e. $\gamma(E)=\gamma_{Born}(E)$, 
see Sec.~\ref{sec:lyapunov} below.

\subsubsection{Applying Berezinskii's approach to wave packets}
\label{subsec:berezinskiiwp}

We now turn towards the generalization of Berezinskii's method to the case of wave packets. Similarly to Eq.~(\ref{bereft}), we can express the asymptotic average particle density in terms of Green functions as follows: 
\begin{eqnarray}
\overline{n(x)} & = & \lim_{\omega\to 0}\frac{\omega}{2\pi i} \int_{-\infty}^\infty {\rm d}E \int_{-\infty}^\infty {\rm d}x''{\rm d}x''' \langle x''|\psi_0\rangle\langle\psi_0|x'''\rangle\nonumber\\
& & \times  \overline{G^{(+)}(x-x'',E+\omega)G^{(-)}(x'''-x,E)}.
\label{eq:GGwp}
\end{eqnarray}
As compared to Eq.~(\ref{bereft}), this expression exhibits additional integrals over the energy $E$ and the two source points $x''$ and $x'''$. Moreover, the average product of Green functions now contains two different source points $x''$ and $x'''$ instead of a single source point $x'$ in Eq.~(\ref{bereft}).

Let us first assume, for simplicity, that no vertices (i.e. no scattering events by the random potential) occur in the region between the two source points $x''$ and $x'''$. An example of a diagram describing such a process is shown in Fig.~\ref{fig:berezinskii}(b). Obviously, it is almost identical to the corresponding Berezinskii diagram in  Fig.~\ref{fig:berezinskii}(a). However, the free propagators $G_0^{(+)}(x'-x_3,E+\omega)$ and $G_0^{(-)}(x'-x_4,E)$ describing the initial propagation from the source $x'$ to the first scattering event at $x_3$ (for the solid line) or $x_4$ (for the dashed line)  in Fig.~\ref{fig:berezinskii}(a), must now be replaced by $G_0^{(+)}(x''-x_3,E+\omega)$ and $G_0^{(-)}(x'''-x_4,E)$  in Fig.~\ref{fig:berezinskii}(b), respectively. Taking into account the explicit form:
\begin{equation}
G_0^{(\pm)}(x_1-x_2,E)=\mp \frac{i}{2 p_E} e^{\pm i p_E|x_1-x_2|},
\label{eq:G0}
\end{equation}
of the free-particle Green functions, with $p_E=\sqrt{E}$, the shifts of the initial positions from $x'$ to $x''$ and from $x'$ to $x'''$, respectively, lead, in total, to an exponential factor: 
\begin{equation}
e^{ip_E (x'''-x'')}=\frac{G_0^{(+)}(x''-x_3,E)G_0^{(-)}(x'''-x_4,E)}{G_0^{(+)}(x'-x_3,E)G_0^{(-)}(x'-x_4,E)},
\label{eq:factor}
\end{equation}
for $x''<x'<x'''<x_{3,4}$ (where the limit $\omega\to 0$ was taken). In other words, the contribution of diagram Fig.~\ref{fig:berezinskii}(b) is given by the contribution of diagram Fig.~\ref{fig:berezinskii}(a) times the above factor, Eq.~(\ref{eq:factor}). 
The same factor applies if $x''$ and $x'''$ are exchanged, i.e. for diagrams in which $x'''<x'<x''<x_{3,4}$, whereas the complex conjugate factor $e^{-ip_E (x'''-x'')}$ is obtained in the cases where 
the particle initially propagates to the left-hand side, i.e. $x_{3,4}<x''<x'<x'''$ or $x_{34}<x'''<x'<x''$.

Inserting these relations into Eq.~(\ref{eq:GGwp}) and using Eq.~(\ref{bereft}), we see that: 
\begin{eqnarray}
\overline{n(x)} & = & \int_{-\infty}^\infty {\rm d}E \int_{-\infty}^\infty {\rm d}x''{\rm d}x'''  \overline{n_E\left(x-\frac{x''+x'''}{2}\right)}
\nonumber\\
&  & \hspace{-1cm} \times \langle x''|\psi_0\rangle\langle\psi_0|x'''\rangle\underbrace{\frac{e^{ip_E (x'''-x'')}+e^{-ip_E (x'''-x'')}}{4\pi p_E}}_{=-\Im\left\{G_0^{(+)}(x'''-x'',E)\right\}/\pi},
\label{eq:berwp1}
\end{eqnarray}
where a factor $1/(2 \pi p_E)$ results from the density of states $\rho_E\simeq 1/(2 \pi p_E)$ (approximated as the density of states for a free particle in 1D)  
in Eq.~(\ref{bereft}), and another factor $1/2$ from the distinction between diagrams where the particle initially propagates to the left or to the right-hand side, respectively, which give exactly the same
contribution $\overline{n_E(x-x')}/2$ (with $x'=\frac{x''+x'''}{2}$) to the total density. 

We recognize the last term of  Eq.~(\ref{eq:berwp1}) as  the negative imaginary part of the free-particle Green function, Eq.~(\ref{eq:G0}), dividided by $\pi$, which, in turn, is related to the spectral function:
\begin{eqnarray}
A_0(p,E) & = & -\frac{1}{\pi}\int_{-\infty}^\infty {\rm d}x''~e^{-ip(x''-x''')}\Im{\{G_0^{(\pm)}(x''-x''',E)}\}\nonumber\\
& = & \delta(E-p^2),
\end{eqnarray} 
of a free particle. Together with the definition of the initial state's Wigner function:
\begin{eqnarray}
W(q,p) & = & \int_{-\infty}^\infty {\rm d}x\ \frac{e^{ipx}}{2\pi}\left<\left.q-\frac{x}{2}\right|\psi_0\right>\left<\psi_0\left|q+\frac{x}{2}\right.\right>, 
\label{eq:wigner}\\
& = & \frac{\pi}{\sqrt{2}}e^{-a^2 p^2-q^2/a^2},
\label{eq:wigner2}
\end{eqnarray}
Eq.~(\ref{eq:berwp1}) turns into:
\begin{equation}
\label{nxafree}
\overline{n(x)}=\int_{-\infty}^\infty {\rm d}E\int_{-\infty}^\infty {\rm d}p{\rm d}q\ W(q,p) A_0(p,E)\overline{n_E(x-q)}.
\end{equation} 
This reproduces Eq.~(\ref{density}) in the limit $L\to\infty$ (i.e. for an infinite system) -- apart from the fact that the average spectral function $A(p,E)$ 
 in the presence of the random potential is approximated by the spectral function $A_0(p,E)$ of the free particle. Remember, however, that our above derivation neglects the presence of the random potential in the vicinity of the source! The remaining diagrams, i.e. those which also contain vertices between $x''$ and $x'''$ are treated in Appendix~\ref{sec:diagrammaticderivation}. As shown there -- under the assumption of a spatially confined initial wave packet (with width $a$ much smaller than the localization length) -- the free particle Green function $G_0^{(\pm)}(x''-x''',E)$ is replaced by the average 
 Green function $\overline{G^{(\pm)}(x''-x''',E)}$ in Eq.~(\ref{eq:berwp1}), and correspondingly $A_0(p,E)$ by $A(p,E)$ in Eq.~(\ref{nxafree}). This concludes the derivation of Eq.~(\ref{density}).

\subsection{Lyapunov exponent}
\label{sec:lyapunov}

As evident from Eq.~(\ref{asymdefeigenfunctionsberezinskii}), the asymptotic state $\overline{n_E(x-x')}$ at fixed energy is determined by the behaviour of eigenfunctions $|\phi_n\rangle$ with energy $E_n=E$. For a one dimensional random potential without 
%tw long-range correlations 
specifically designed correlation function \cite{Izrailev}, all eigenfunctions are exponentially localized, and the rate of exponential decay defines the Lyapunov exponent:
\begin{equation}
\gamma(E)=-\lim_{x\to\infty} \overline{\ln\left(||\psi_E(x)||\right)}/|x|.
\label{eq:lyapunovdef}
\end{equation}
%tw minus sign added!
Its inverse $\xi(E)=1/\gamma(E)$ is called localization length. Since the 
%tw wavefunction intensity in the localized regime exhibits a log-normal distribution, 
average of the logarithm in Eq.~(\ref{eq:lyapunovdef}) differs from the logarithm of the average,
the average intensity $\overline{||\psi_E(x)||^2}$ 
%tw -- which must be distinguished from the average of the logarithm in Eq.~(\ref{eq:lyapunovdef}) -- can  be shown to 
decays with a different exponent, which can be shown to be given by 
$\gamma(E)/2$  
%tw \cite{Mueller}.
\cite{Izrailev}.
This agrees with the behaviour of $\overline{n_E(x)}$, see Eq.~(\ref{eigenlocstate}), for large $x$ \cite{Gogolin2}:
\begin{equation}
\overline{n_E(x)}\underset{x\to\infty}{\longrightarrow} x^{-3/2}e^{-\gamma(E)|x|/2}.
\end{equation}
As already mentioned above, the diagrammatic approach of Berezinskii and Gogolin 
yields the Born approximation  of the Lyapunov exponent \cite{Lugan}:
\begin{eqnarray}
\gamma_{{\rm Born}}(E) & = & \frac{1}{8E}\int_{-\infty}^{\infty}dx\ C_2(x)\cos{(2 p_E x)} \label{bornlyapunov}\\
& = & \sqrt{\frac{\pi}{2}} \frac{\sigma_c V_0^2}{4 E}e^{-2 \sigma_c^2 E}, \label{bornlyapunov2}
\end{eqnarray}
where we have used our explicit form of the correlation function $C_2(x)$ as given by Eq.~(\ref{correlationx}) in Eq.~(\ref{bornlyapunov2}).
The Born approximation is valid at high energies, i.e. for $p_E\gg \gamma_E$. For weak random potentials, i.e. if $V_0\sigma_c^2\ll 1$ 
%tw -- which we will assume in the following -- 
as explained in Sec.~\ref{sec:model}, %tw
Eq.~(\ref{bornlyapunov2}) is therefore valid if  $E\gg (\sigma_c V_0^2)^{2/3}$.
For lower energies, however, the perturbative expression given by Eq.~(\ref{bornlyapunov}) fails, as evident from the divergent prefactor $1/E$ as $E\to 0$. Since our wave packet, initially given by Eq.~(\ref{initialgaussianstate}), also exhibits contributions with low energies (see Sec.~\ref{subsec:spectraldensity}, where the energy distribution of the wave packet will be determined), we present in the following a more precise analytical evaluation of the Lyapunov exponent, which can be applied also  for small  energies. We then use this expression for $\gamma(E)$ in the formula of Gogolin, Eq.~(\ref{eigenlocstate}).

For this purpose, we rely on the work of Halperin \cite{Halperin} and Thouless \cite{Thouless,Thouless2}, who have found an exact analytical result  for the Lyapunov exponent in a one-dimensional uncorrelated potential (also called \lq white noise\rq\ potential), i.e. a potential with correlation function $C_2(x)=D\delta(x)/2$, where $D$ indicates the strength of the white noise. First, Halperin derived the following expression for the average cumulative density of states, defined as the integral $N(E)=\int_{-\infty}^E {\rm d}E'~\rho(E')$ over the average density of states $\rho(E)$, Eq.~(\ref{eq:rho}):
\begin{equation}
\label{halpnumber}
N(E)=\frac{(D/4)^{1/3}}{\pi^2 \left\{{\rm Ai}^2\left[-E\left(\frac{16}{D^2}\right)^{1/3}\right]+{\rm Bi}^2\left[-E\left(\frac{16}{D^2}\right)^{1/3}\right]\right\}},
\end{equation}
where ${\rm Ai}$ and ${\rm Bi}$ denote Airy functions of the first and second kind, respectively. Second, Thouless found the following 
%tw
exact %tw
analytical relation between the Lyapunov exponent and the cumulative density of states:
\begin{equation}
\label{gamathoules}
\gamma(E)-\gamma_0(E)={\mathcal P}\displaystyle\int_{-\infty}^{\infty}dE'\frac{N(E')-N_0(E')}{E'-E},
\end{equation}
where ${\mathcal P}$ denotes the principal value, and $N_0(E)=\Re\{\sqrt{E}\}/\pi$ the cumulative density of states of a free particle. Furthermore,
$\gamma_0(E)=\Im\{\sqrt{E}\}$ represents the Lyapunov exponent for a free particle. It vanishes for positive energies, whereas
$\gamma_0(E)=\sqrt{-E}$ for $E<0$. 
Eq.~(\ref{gamathoules}) can be read as a Kramers-Kronig relation between the quantities $N(E)$ and $\gamma(E)/\pi$. In other words, if there exists a complex function $F(z)$ which is analytic in the upper half of the complex plane and tends to $i\sqrt{z}/\pi$ as $|z|\to\infty$, and if the imaginary part of this function coincides with  the cumulative density of states, i.e.,
$N(E)=\Im \{F(E)\}$, then the Lyapunov exponent is given by $\pi$ times the real part of this function, i.e., $\gamma(E)=\pi \Re \{F(E)\}$.
Such a function indeed exists \cite{Thouless2}:
\begin{equation}
F(E)=\frac{D^{1/3}}{4^{1/3}\pi} \frac{{\rm Ai}'\left[-E\left(\frac{16}{D^2}\right)^{1/3} \right]+ i {\rm Bi}'\left[-E\left(\frac{16}{D^2}\right)^{1/3}\right]}{{\rm Ai}\left[-E\left(\frac{16}{D^2}\right)^{1/3}\right]+i {\rm Bi}\left[-E\left(\frac{16}{D^2}\right)^{1/3}\right]},
\end{equation}
which yields the Lyapunov exponent:
\begin{equation}
\label{lyapunovwhite}
\gamma(E)=-\left(\frac{D}{4}\right)^{1/3}\frac{M'\left[-E\left(\frac{16}{D^2}\right)^{1/3}\right]}{M\left[-E\left(\frac{16}{D^2}\right)^{1/3}\right]},
\end{equation}
where $M(y)=\sqrt{{\rm Ai}^2(y)+{\rm Bi}^2(y)}$.

Eq.~(\ref{lyapunovwhite})
 is exact for the white noise potential. Our idea is now to choose the strength $D$ of the white noise in such a way that, for a given energy $E$, the Lyapunov exponent of the white noise potential agrees with the one of our correlated potential. For this purpose, we first look at the asymptotic bevaviour of $\gamma(E)$, given by Eq.~(\ref{lyapunovwhite}), for large energies:
\begin{equation}
\gamma(E)\underset{E\to\infty}{\longrightarrow}\frac{D}{16 E}.
\end{equation}
In order to reproduce the result of the Born approximation, 
we therefore choose
$D=16 E \gamma_{\rm Born}(E)$, or, taking into account Eqs.~(\ref{bornlyapunov},\ref{bornlyapunov2}):
\begin{eqnarray}
D &  = & 2\int_{-\infty}^{\infty}dx\ C_2(x)\cos{(2 p_E x)}\label{eq:DE}\\
& = & \sqrt{\frac{\pi}{2}} 4\sigma_c V_0^2 e^{-2 \sigma_c^2 E}. \label{eq:DE2}
\end{eqnarray}
As already mentioned above, the Born approximation is valid for large energies $E\gg E^{(1)}_{\rm min}$, where
$E^{(1)}_{\rm min}=(\sigma_c V_0^2)^{2/3}$.
On the other hand, we can also define a regime of low energies $E\ll E^{(2)}_{\rm max}$ where $E^{(2)}_{\rm max}=\sigma_c^{-2}$. This corresponds to the condition that the wave length $\lambda_E=2\pi/p_E$ is much larger than the correlation length $\sigma_c$.
In this regime, the wave is not able to resolve the correlations of the potential, which can therefore be approximated by a white noise potential $C_2(x)=D\delta(x)/2$, with $D/2=\int_{-\infty}^\infty {\rm d}x~C_2(x)$.  This again agrees with Eq.~(\ref{eq:DE}), where we may approximate
$\cos(2 p_E x)\simeq 1$ for low energies. Thus, Eq.~(\ref{eq:DE}) together with  Eq.~(\ref{lyapunovwhite}) yields a good approximation for the Lyapunov exponent, both, for high and for low energies. Moreover, remember that we assumed $V_0\sigma_c^2\ll 1$, which is equivalent to $E^{(1)}_{\rm min}\ll E^{(2)}_{\rm max}$. Since this implies that the two above regimes overlap, our approximation is expected to hold in the entire range of energies.

\begin{figure}[h]
\includegraphics[width=8.5cm]{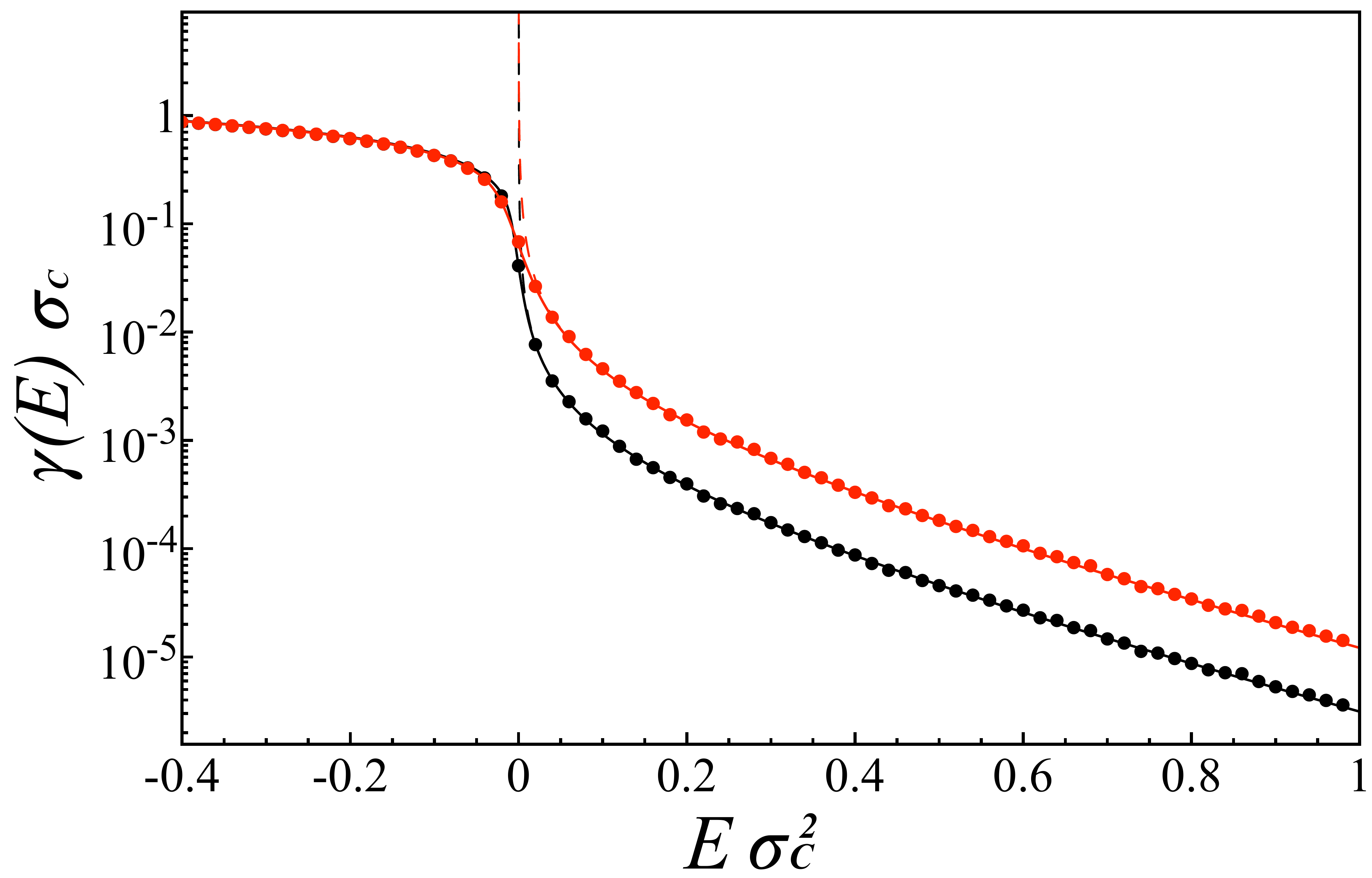}
\caption{\label{coeff} Lyapunov exponent (in units of $\sigma_c^{-1}$) as a function  of energy (in units of $\sigma_c^{-2}$) for two different strengths of the random potential, $V_0\sigma_c^2=0.0165$ (black) and $V_0\sigma_c^2=0.0325$ (red), both of which fulfill the condition $V_0\sigma_c^2\ll 1$ of a weak random potential.
Our analytical prediction (solid lines), based on the exact expression for a white noise potential, Eq.~(\ref{lyapunovwhite}), with noise strength $D$ chosen according to Eq.~(\ref{bornlyapunov}), agrees well with the result of numerical transfer matrix calculations (dots) in the entire range of energies, and for both values of the potential.
In contrast, the Born approximation, Eq.~(\ref{bornlyapunov}) (dashed lines) exhibits a divergence at $E\to 0$.
}
\end{figure}

This expectation is confirmed by numerical calculations of the Lyapunov exponent using the transfer matrix method \cite{Alberto}, see Fig.~\ref{coeff}. We see that our above analytical expression, Eq.~(\ref{lyapunovwhite}) together with Eq.~(\ref{eq:DE2}) (solid lines), agrees well with the numerical results (dots) in the entire range of energies, in particular also for negative energies. In contrast, the Born approximation (dashed lines) exhibits a divergence at low energies.

\subsection{Spectral function and energy distribution}
\label{subsec:spectraldensity}

The last element needed in order to complete our theoretical description of the  average asymptotic density $\overline{n(x)}$ based on 
Eq.~(\ref{density}) is the spectral function $A(p,E)$.
 In general, it is defined as the negative imaginary part of the average Green function 
 %tw
 (represented in momentum space) %tw
 divided by $\pi$:
\begin{equation}
A(p,E)=-\frac{1}{\pi}\Im{\{\overline{\tilde{G}^{(+)}(p,E)}\}}.
\label{eq:ApE}
\end{equation}
The spectral function describes the relation between energy and momentum in the presence of the random potential. More precisely, it
defines the probability that a state with momentum $p$ has energy $E$.

In order to calculate the average Green function, we use the Dyson equation \cite{Jorgen}
\begin{equation}
\overline{\tilde{G}^{(+)}(p,E)}=\tilde{G}_0^{(+)}(p,E)\left(1+\tilde{\Sigma}^{(+)}(p,E)\overline{\tilde{G}^{(+)}(p,E)}\right)\label{eq:dyson}
\end{equation}
together with a suitable approximation for the self-energy $\tilde{\Sigma}^{(+)}(p,E)$. Here:
\begin{eqnarray}
\tilde{G}_0^{(+)}(p,E) & = & \int_{-\infty}^\infty {\rm d}x~e^{i p x} G_0^{(+)}(x,E),\nonumber\\
& = & \frac{1}{E-p^2+i\eta},
\end{eqnarray}
with infinitesimal positive quantity $\eta$, denotes the Fourier transform of the free-particle Green function, Eq.~(\ref{eq:G0}). The solution of Eq.~(\ref{eq:dyson}) reads:
\begin{equation}
\overline{\tilde{G}^{(+)}(p,E)}=\frac{1}{E-p^2-\tilde{\Sigma}^{(+)}(p,E)}.\label{eq:Gtilde}
\end{equation}
Assuming a weak random potential (and, correspondingly, small $\Sigma$), we see that the main contribution to $\overline{\tilde{G}^{(+)}(p,E)}$ originates from momenta $p\simeq p_E=\sqrt{E}$, where the term $E-p^2$ in the denominator of Eq.~(\ref{eq:Gtilde}) vanishes.
We hence neglect the momentum dependence of the self energy, i.e. we replace Eq.~(\ref{eq:Gtilde}) by:
\begin{equation}
\overline{\tilde{G}^{(+)}(p,E)}=\frac{1}{E-p^2-\Sigma^{(+)}(E)},\label{eq:Gtilde2}
\end{equation}
where:
\begin{equation}
\Sigma^{(+)}(E)= \tilde{\Sigma}^{(+)}(p_E,E).
\end{equation}
The average Green function then takes the following form in position space:
\begin{equation}
\overline{G^{(+)}(x_1-x_2,E)}=-\frac{i}{\tilde{p}_E} e^{i \tilde{p}_E|x_1-x_2|},
\label{eq:Gposition}
\end{equation}
which is similar to the free-particle Green function, Eq.~(\ref{eq:G0}), but with complex effective wave vector
\begin{equation}
\tilde{p}_E=\sqrt{E-\Sigma^{(+)}(E)}
\label{eq:ptilde}
\end{equation}
instead of $p_E$. 

The simplest approximation for $\Sigma^{(+)}(E)$ is given by the Born approximation, which amounts to an expansion up to second order in the potential $V$, i.e. $\Sigma^{(+)}_{\rm Born}(E)=\overline{VG_0^{(+)}(E)V}$, or:
\begin{eqnarray}
\label{borna}
\Sigma^{(+)}_{\rm Born}(E) &=&\int_{-\infty}^{\infty} {\rm d}x\ C_2(x)\ e^{i p_E x}G_0^{(+)}(x,E),\label{eq:Sigmaborn1}\\
&  = &  \frac{d(E)}{i \sqrt{E}},
\label{eq:Sigmaborn2}
\end{eqnarray}
where: 
\begin{equation}
d(E)=\frac{\pi\sigma_c V_0^2}{4}\left[1+e^{-2 E\sigma_c^2}\left(1+i~{\rm erfi}\left(\sqrt{2 E}\sigma_c\right)\right)\right],
\end{equation}
with imaginary error function ${\rm erfi}(z)$. Similarly as in the case of the Lyapunov exponent, Eq.~(\ref{bornlyapunov}), the Born approximation 
is valid only for high energies, and a divergence proportional to $1/\sqrt{E}$ is observed for $E\to 0$, see Eq.~(\ref{eq:Sigmaborn2}).

A more precise estimation is obtained by the self-consistent Born approximation, where the free-particle Green function $G_0^{(+)}(E)$ in Eq.~(\ref{eq:Sigmaborn1}) is replaced by the average Green function $\overline{G^{(+)}(E)}$, Eq.~(\ref{eq:Gposition}), which, in turn, depends on the self energy, see Eq.~(\ref{eq:ptilde}).  The resulting self-consistent equation for $\Sigma(E)$, however, cannot be analytically solved. Since, as pointed out above, the failure of the Born approximation at low energies arises from the 
denominator $1/p_E=1/\sqrt{E}$ in Eq.~(\ref{eq:Sigmaborn2}), our approach in order to obtain an analytical result is to treat only this denominator in a self-consistent way. We therefore replace $1/p_E\to1/\tilde{p}_E$, i.e. $1/\sqrt{E}\to 1/\sqrt{E-\Sigma^{(+)}(E)}$ in the denominator of Eq.~(\ref{eq:Sigmaborn2}), and obtain the following self-consistent equation for $\Sigma$:
\begin{equation}
\Sigma^{(+)}(E)=\frac{d(E)}{i\sqrt{E-\Sigma^{(+)}(E)}}.
\end{equation}
This equation has the following  unique solution with negative imaginary part:
\begin{widetext}
\begin{equation}
\Sigma^{(+)}(E)=\left[d(E)\right]^{2/3} \left(\frac{\epsilon}{3} + \frac{(-1)^{2/3}\epsilon^2}{3\left(\epsilon^3+\frac{3}{2}\left(9+\sqrt{81+12\epsilon^3}\right)\right)^{1/3}}-
\frac{1}{3}(-1)^{1/3} (\epsilon^3+\frac{3}{2}\left(9+\sqrt{81+12\epsilon^3}\right)^{1/3}\right),
\label{eq:Sigmafinal}
\end{equation}
\end{widetext}
where: 
\begin{equation}
\epsilon=\frac{E}{\left[d(E)\right]^{2/3}}.
\end{equation}
Together with Eqs.~(\ref{eq:ApE}) and (\ref{eq:Gtilde2}), this concludes our analytic calculation of the spectral function.

Finally, the energy distribution $P(E)$ of our wave packet is extracted from the spectral function according to Eq.~(\ref{PE}).
Using the Wigner function $W(q,p)$ of the initial state given in Eq.~(\ref{eq:wigner2}), the integral over $q$ and $p$ can be performed analytically:
\begin{equation}
P(E)=\frac{a}{\sqrt{\pi}}\Im\left\{\frac{e^{a^2\left(\Sigma^{(+)}(E)-E\right)}{\rm erfc}\left(a\sqrt{\Sigma^{(+)}(E)-E}\right)}{\sqrt{\Sigma^{(+)}(E)-E}}\right\},
\label{eq:PEfinal}
\end{equation}
with complementary error function ${\rm erfc}(z)$. As we have checked, this expression, with $\Sigma^{(+)}(E)$ given by Eq.~(\ref{eq:Sigmafinal}), 
yields a normalized distribution, i.e. $\int_{-\infty}^\infty{\rm d}E~P(E)=1$, which underlines the consistency of our approach.
Its quality in terms of agreement with numerical data will be discussed in Sec.~\ref{subsec:result_PE}. There, we will also compare our prediction $P(E)$ with the energy distribution of a free particle, using the spectral function $A_0(p,E)=\delta(E-p^2)$ instead of $A(p,E)$:
\begin{equation}
P_0(E)=\frac{a}{\sqrt{2 \pi E}}e^{-a^2 E/2}.
\label{eq:P0}
\end{equation}

We summarize our theoretical approach presented in this chapter as follows:
First, Eq.~(\ref{densitysimple}) describes the asymptotic average density $\overline{n(x)}$
as the integral of the density $\overline{n_E(x)}$ at fixed energy, weighted with the
 energy distribution $P(E)$. The former is obtained using Gogolin's formula, Eq.~(\ref{eigenlocstate}), where we insert the Lyapunov exponent $\gamma(E)$ defined by Eqs.~(\ref{lyapunovwhite}) and (\ref{eq:DE}). Finally, the energy distribution $P(E)$ is given by Eqs.~(\ref{eq:Sigmafinal}) and (\ref{eq:PEfinal}), as described above.

\section{Numerical results}
\label{sec:results}

In this section, we test the validity of our theory presented in Sec.~\ref{sec:theory} by comparison with numerical results.
The latter are produced by diagonalization of the Hamiltonian $H$, see Eq.~(\ref{eq:H}), i.e. we calculate the spectrum $\{E_n\}$ and corresponding eigenfunctions $\{|\phi_n\rangle\}$ for many different realizations of the random potential $V(x)$.
For the diagonalization, we use a finite-element discrete variable representation (FEDVR) \cite{Iva,TN,Schneider}. Our system of length 
$L=800\sigma_c$ (i.e. much larger than the correlation length $\sigma_c$) is divided into $10^4$ finite elements of size $0.08\sigma_c$ (i.e. much smaller than $\sigma_c$). Each element, in turn, is discretized using 5 basis functions (among them 2 bridge functions which connect this element to the neighbouring ones) \cite{TN,Schneider}. Thereby, the Hamiltonian is represented as a matrix of dimension $M=4\times 10^4$, which can be diagonalized numerically. As already mentioned in Sec.~\ref{sec:model}, we use periodical boundary conditions, i.e. the point $x=-L/2$ is identical to the point $x=L/2$.

\subsection{Energy distribution}
\label{subsec:result_PE}

\begin{figure}[h]
\includegraphics[width=8.5cm]{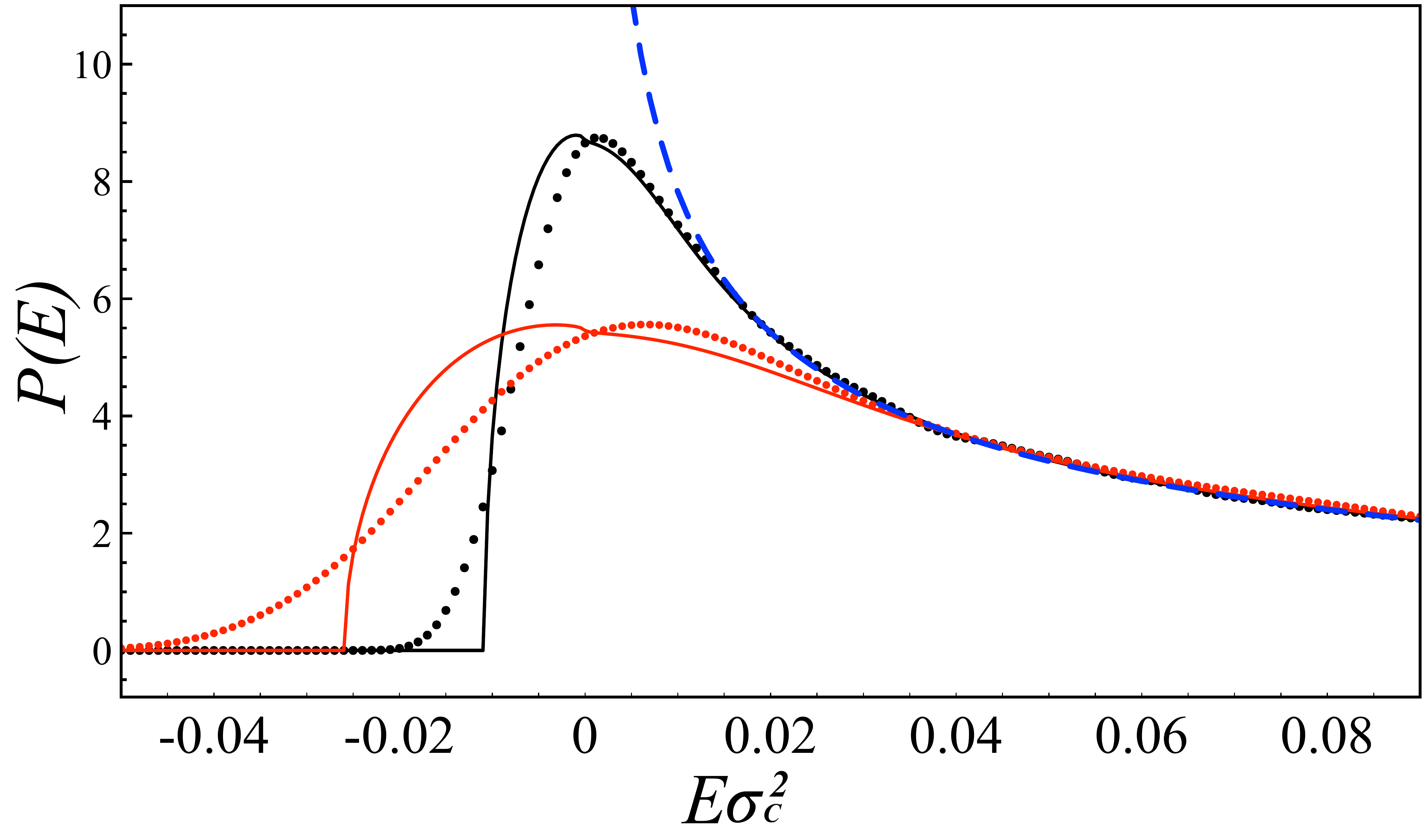}
\caption{\label{P}
Energy probability distribution, $P(E)$, for width $a=\sqrt{2}~\sigma_c$ of the initial wave packet and two different strengths of the random potential, $V_0= 0.0165\sigma_c^{-2}$ (black) 
and  $0.0325\sigma_c^{-2}$ (red). The results of the numerical diagonalization (dotted lines), averaged over $2000$ realizations of the random potential and, in addition, smoothed by an
%tw running average of length $\Delta E=0.01$
exponentially moving average with smoothing constant 0.007, %tw
agree well with our analytical prediction (solid lines) based on the self-consistent Born approximation, Eq.~(\ref{eq:PEfinal}),  for positive energies. For negative energies, the theory exhibits a sharp cutoff which is not present in the numerical data. For large energies ($E>V_0$), all curves converge to the distribution $P_0(E)$ of a free particle (blue dashed line).}
\end{figure}

Let us start with the energy distribution $P(E)$ introduced in Sec.~\ref{subsec:spectraldensity}. From the numerically determined eigenvalues and eigenfunctions of $H$, this distribution is extracted as follows:
\begin{equation}
\label{PEn}
P(E)=\overline{\sum_{n=0}^{M}|\langle \phi_n|\psi_0\rangle|^2\delta(E-E_n)}.
\end{equation}  
Fig.~\ref{P} shows the numerical result (dotted lines) together with our analytical prediction, Eq.~(\ref{eq:PEfinal}) (solid lines), for initial width $a=\sqrt{2}\sigma_c$ and two different values of the potential $V_0=0.0165\sigma_c^{-2}$ (black) and $V_0=0.0325\sigma_c^{-2}$ (red). In addition, we also show the energy distribution of a free particle,
Eq.~(\ref{eq:P0}) (blue dashed line). As expected, all curves agree well for energies larger than $V_0$. For smaller energies, the free-particle distribution exhibits a divergence if
 $E\to 0$. In contrast, our version of the self-consistent Born approximation reproduces well the numerical distribution in the entire range of positive energies. For negative energies, the agreement is less precise, due to the fact that our analytical result exhibits a sharp cutoff at a certain minimum energy, whereas the numerical distribution decays smoothly with decreasing energy. We note that a more accurate description in the range of negative energies can be achieved by a recently developed semiclassical approach \cite{Trappe}. This difference, however, does not significantly affect the shape of the asymptotic state, see Sec.~\ref{subsec:state}, which, according to our theory, depends on $P(E)$ only through the energy dependence of 
 the Lyapunov exponent $\gamma(E)$, see Fig.~\ref{coeff}.
 
\subsection{Asymptotic average density}
\label{subsec:state}

\begin{widetext}
\begin{figure*}
\begin{center}
\includegraphics[width=17cm]{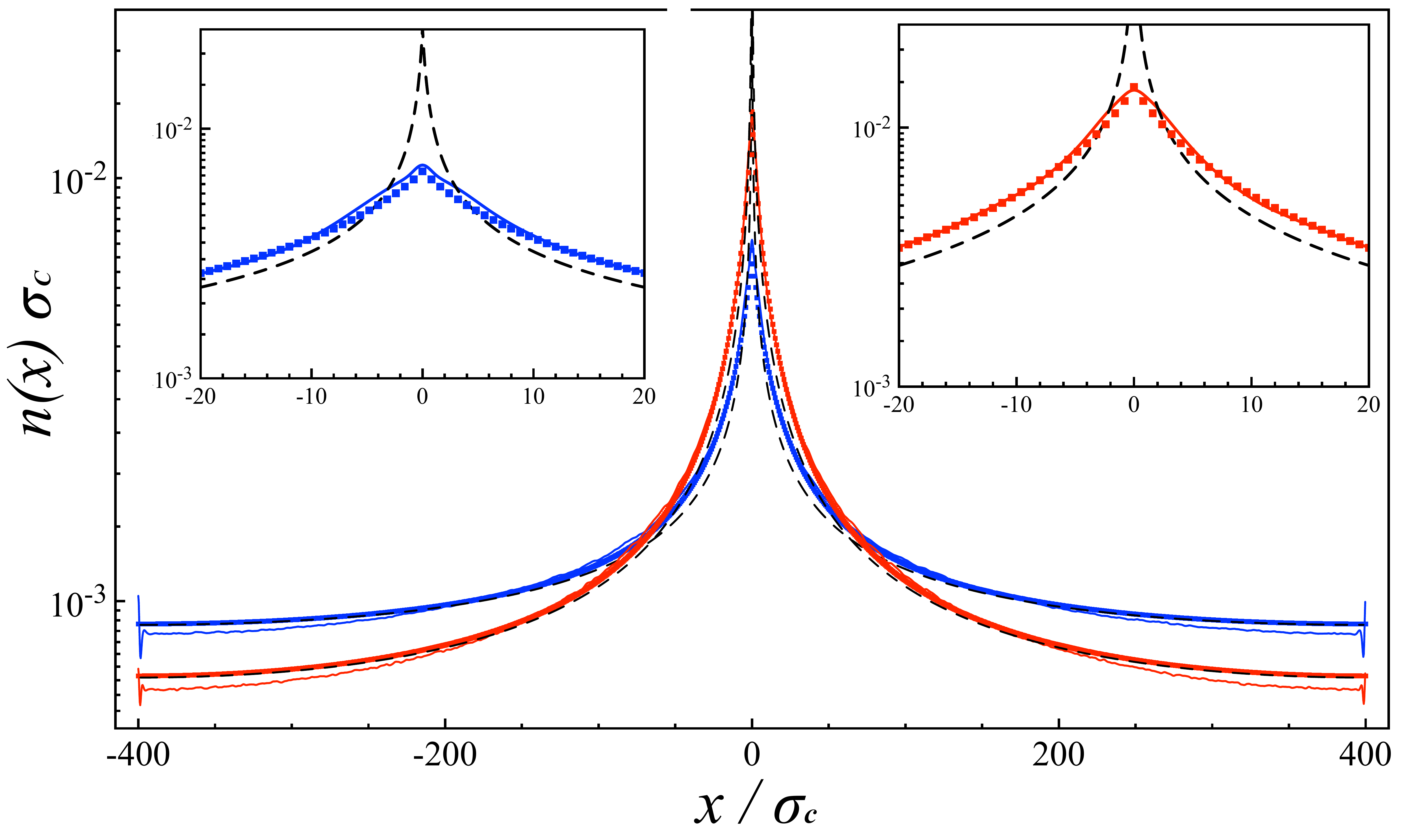}
\caption{\label{A} 
Numerical and analytical results for the asymptotic average density $\overline{n(x)}$, for width $a=\sqrt{2}~\sigma_c$ of the initial wave packet and two different strengths of the random potential, $V_0= 0.0165\sigma_c^{-2}$ (blue) 
and  $0.0325\sigma_c^{-2}$ (red).
The numerical result given by Eq.~(\ref{asymdefeigenfunctions}) (solid lines), averaged over $2000$ realizations of the random potential,
agrees well with our theoretical description, Eq.~(\ref{density}) (squares), in particular also close to the center of the wave packet.
In contrast, a simplified version of the theory (dashed lines) using the spectral distribution $P_0(E)$ of a free particle instead of $P(E)$, and the Born approximation $\gamma_{\rm Born}(E)$ instead of the Lyapunov exponent $\gamma(E)$, predicts an unphysical divergence of the density at $x=0$ (see inset).}
\end{center}
\end{figure*}
\end{widetext}

We now turn towards the main result of the paper concerning the asymptotic average density $\overline{n(x)}$ defined in Sec.~\ref{sec:model}.
Fig.~\ref{A} shows a comparison of our theoretical prediction based on Eq.~(\ref{densitysimple})  with the result extracted from numerically determined eigenfunctions according to Eq.~(\ref{asymdefeigenfunctions}). In addition, we also show the result of a simplified theory (used in \cite{Sanchez}), where the energy distribution $P(E)$ in Eq.~(\ref{densitysimple}) is replaced by the free-particle distribution $P_0(E)$, see Eq.~(\ref{eq:P0}), and the Lyapunov exponent $\gamma(E)$ entering in Eq.~(\ref{eigenlocstate}) is replaced by its Born approximation $\gamma_{\rm Born}(E)$, see Eq.~(\ref{bornlyapunov}). Overall, we see that our theory (solid lines) gives a good description of the numerical data (squares), both, in the wings of the spatial profile and, remarkably, also close to the center of the wave packet (inset). Due to the small-$x$ behaviour of $\overline{n_E(x)}$, which can be deduced from Eq.~(\ref{eigenlocstate})
\begin{equation}
\overline{n_E(x)}\underset{x\to 0}{\longrightarrow}\frac{2\gamma(E)}{3}\Bigl(1-2\gamma(E)x\Bigr)
\label{eq:lineardecay}
\end{equation}
our theory predicts a linear decay of the profile  at small $x$, which can be evaluated by integrating Eq.~(\ref{eq:lineardecay}) over $E$, see Eq.~(\ref{densitysimple}). (Note that only the term $j=0$ gives a significant contribution in the center.) In contrast, the simplified theory agrees with our improved theory in the wings, but fails in the center where it predicts an unphysical divergence as $x\to 0$.

Upon closer inspection, however, we note a small deviation between theory and numerics in the wings of the profile. Furthermore, the numerical data exhibit an interference feature at $x\simeq L/2$ and $x\simeq -L/2$, which is absent in our theory.

\subsection{Agreement theory vs. numerics in different energy intervals}

\begin{widetext}
\begin{figure*}
\begin{center}
\includegraphics[width=5.5cm]{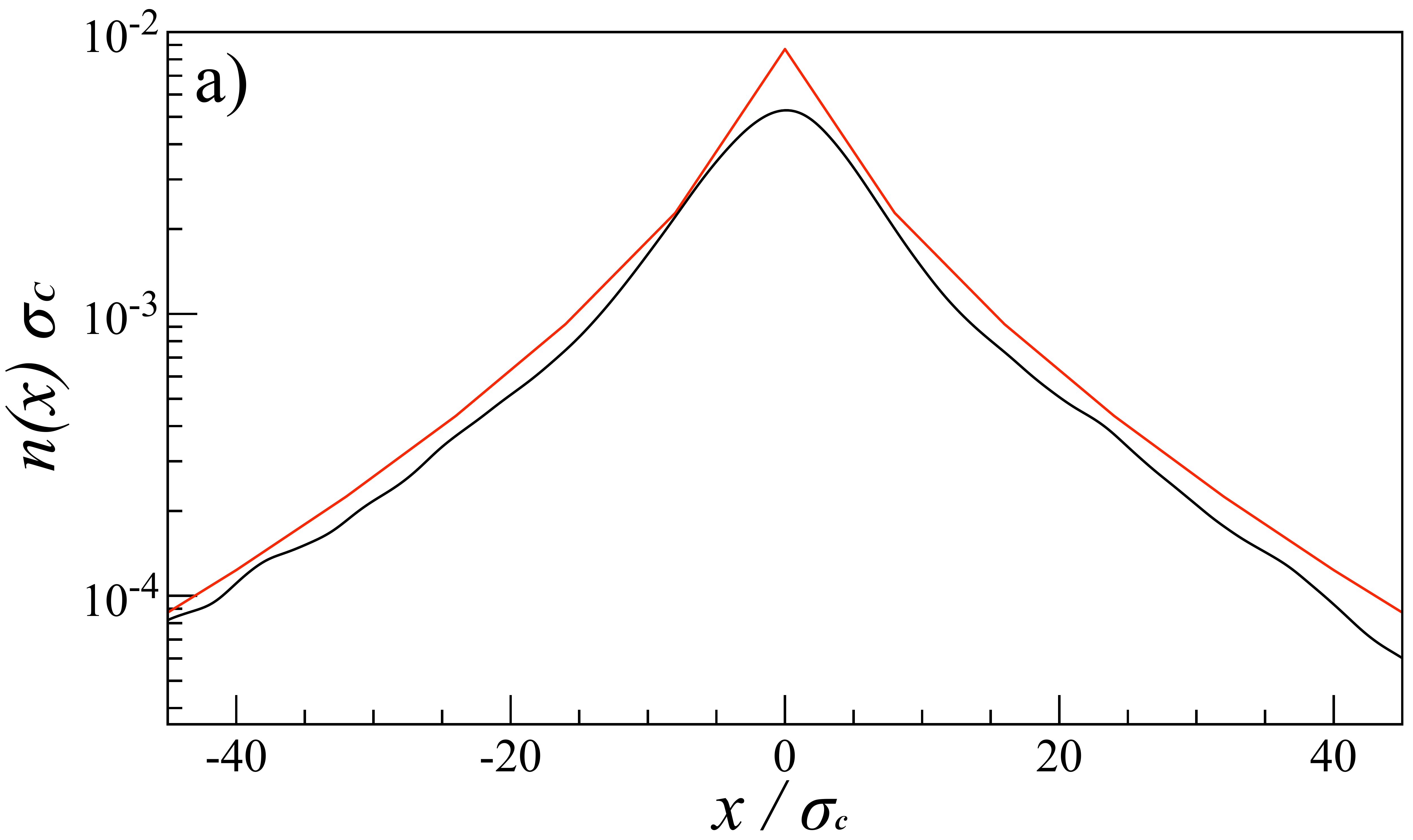}
\includegraphics[width=5.5cm]{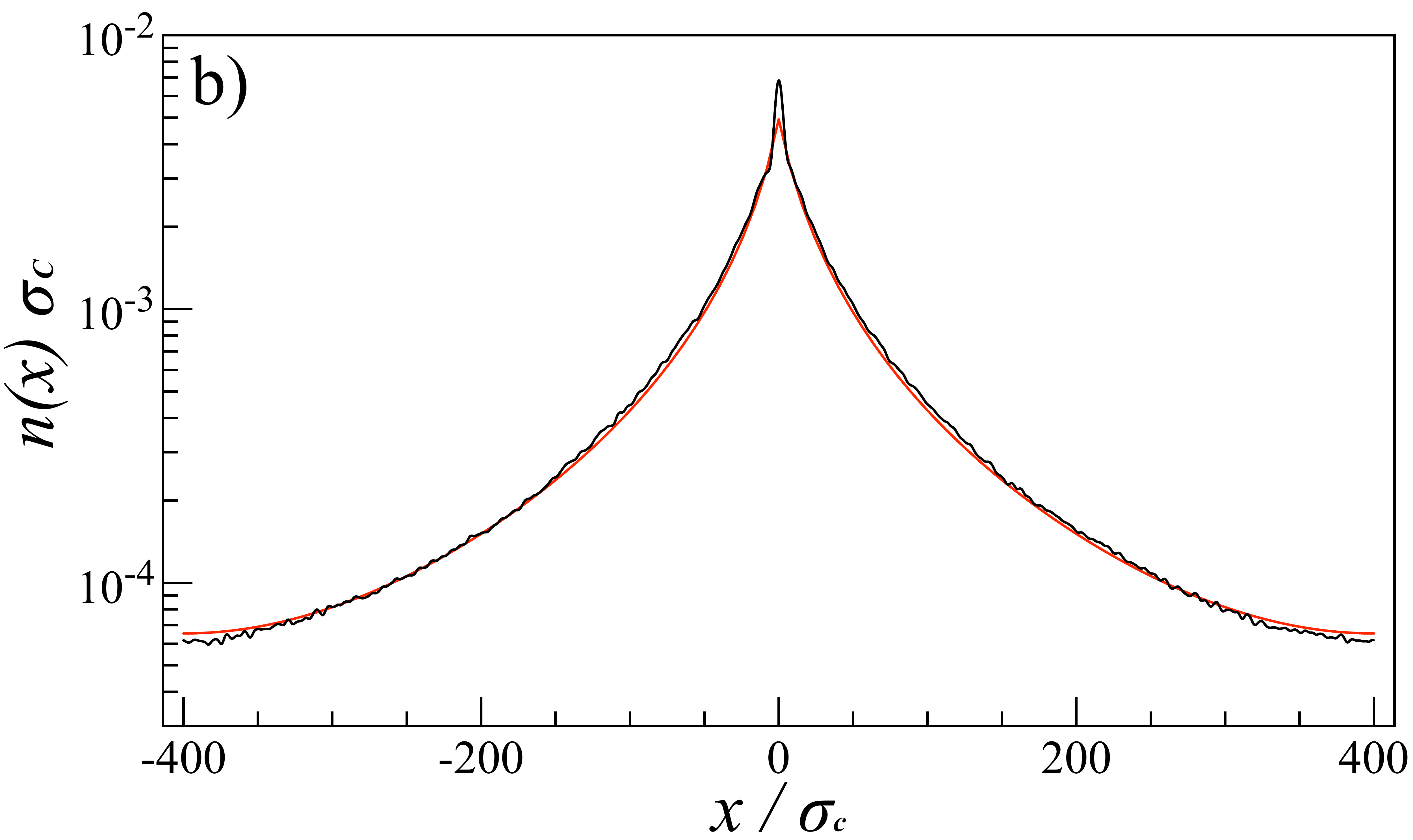}
\includegraphics[width=5.5cm]{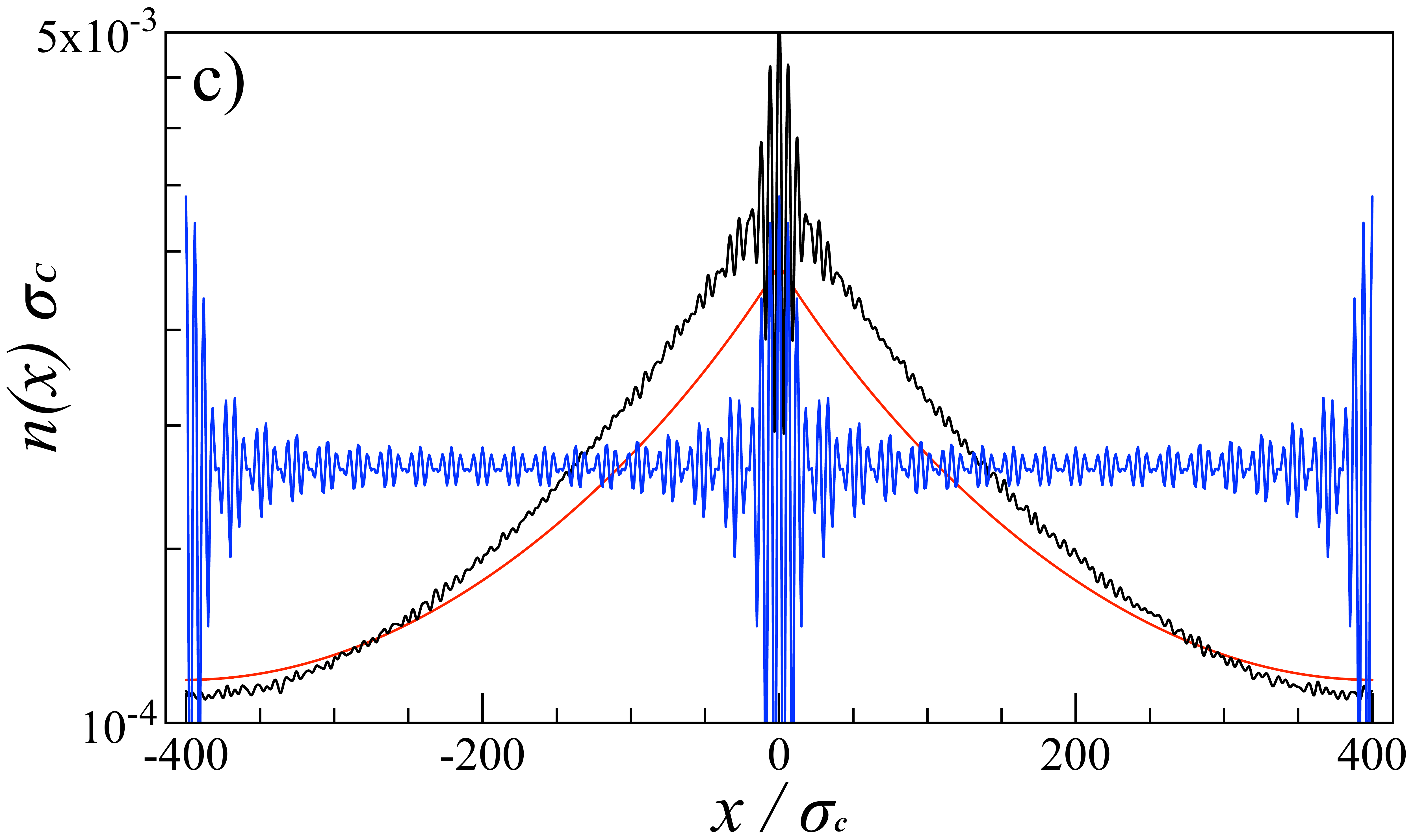}
\includegraphics[width=5.5cm]{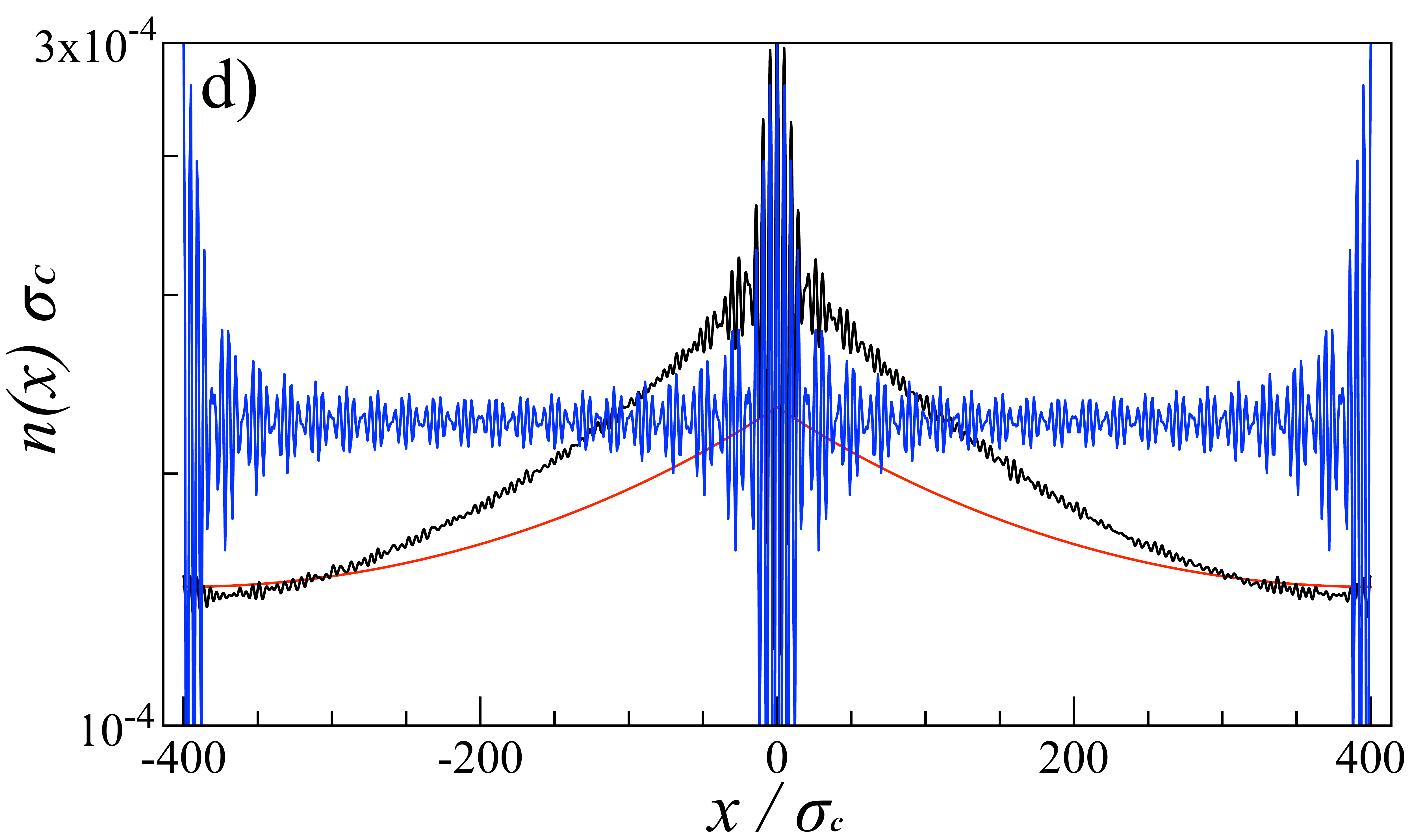}
\includegraphics[width=5.5cm]{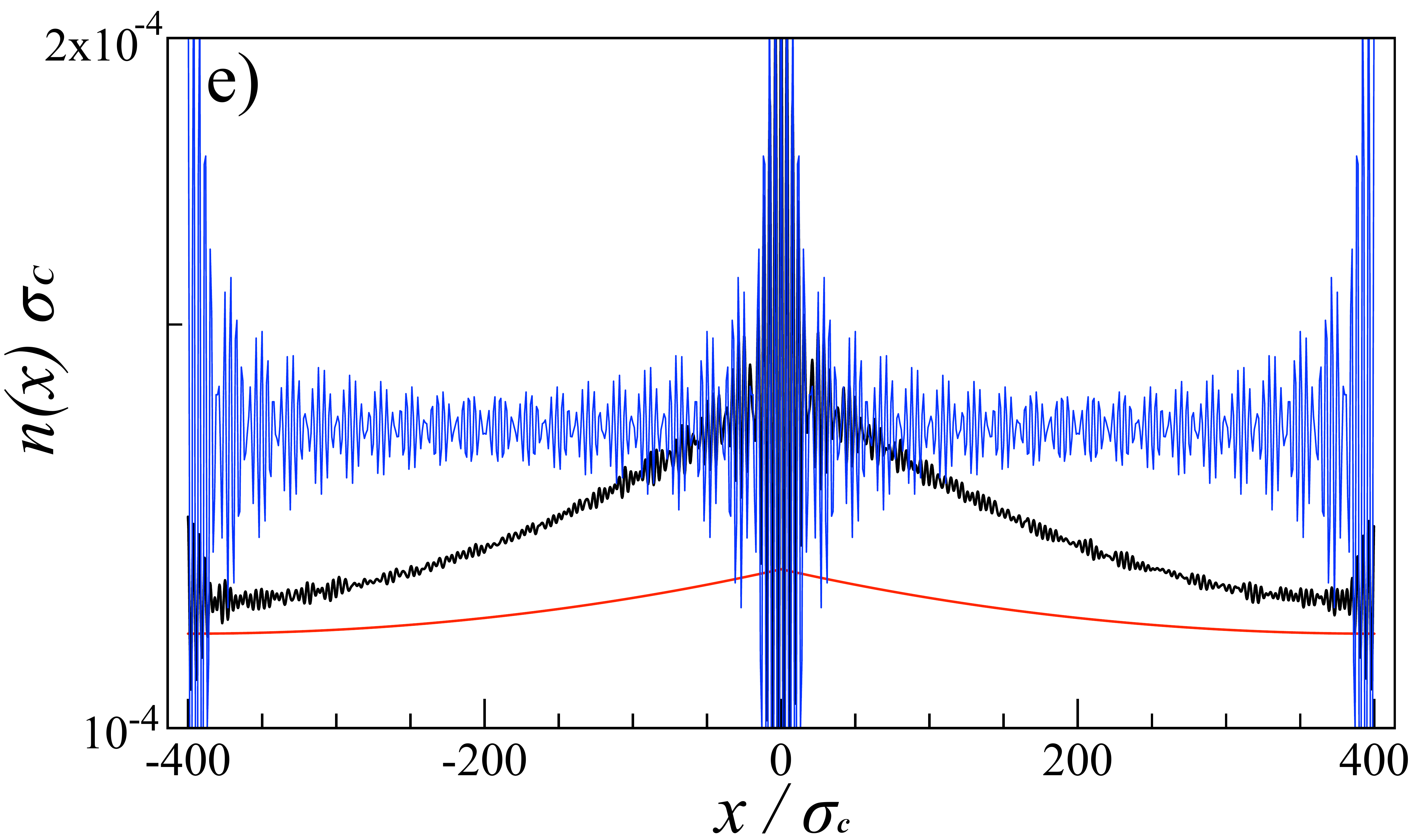}
\includegraphics[width=5.5cm]{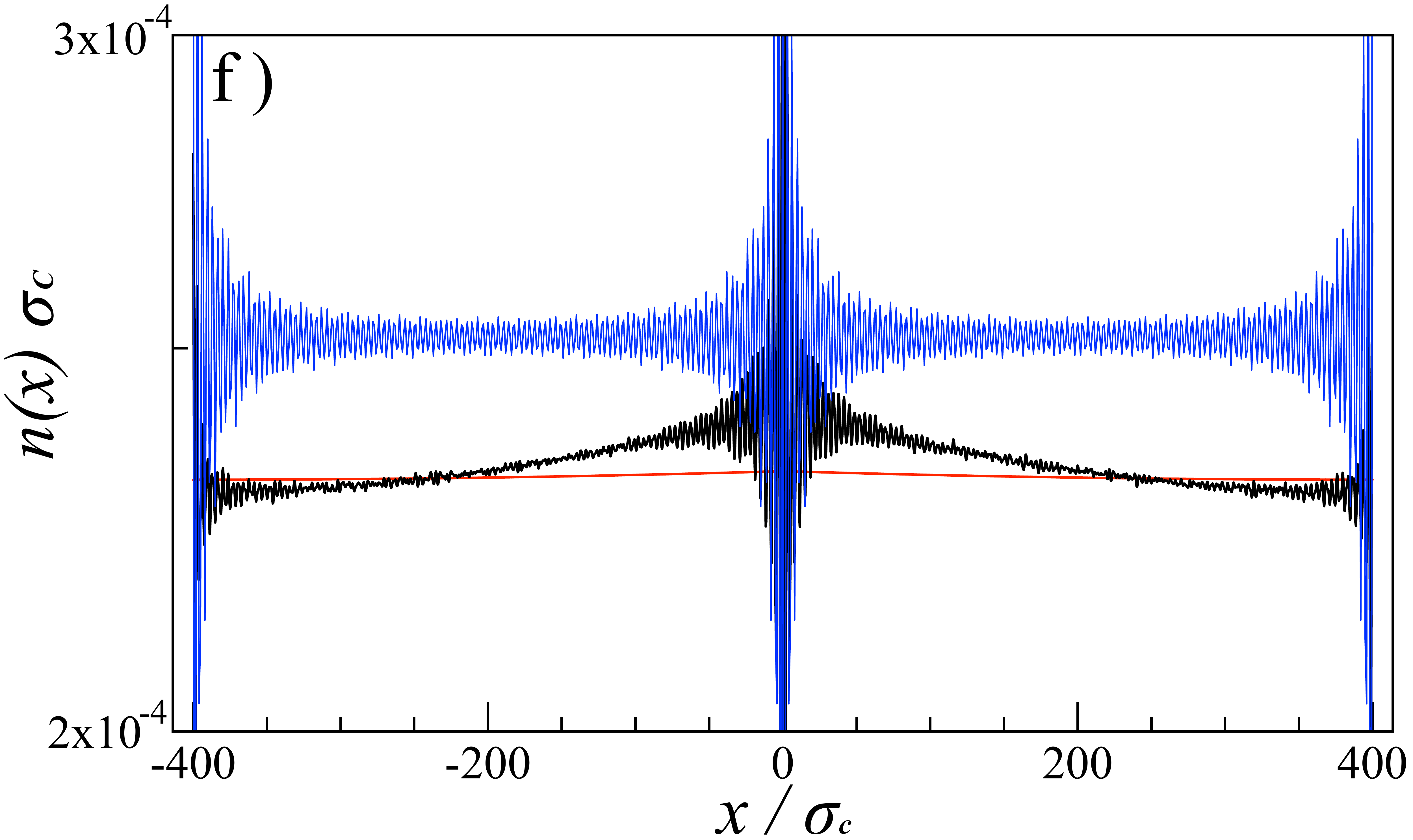}
\caption{\label{ver1}
%tw
Comparison between numerical (black) and theoretical (red) density profiles, see Eqs.~(\ref{nintervalnum}) and (\ref{nintervaltheo}), respectively, in different energy windows a) $(E_1,E_2)=(-\infty,~0)$, b) $(E_2,E_3)=(0,~0.103)$, c) $(E_3,E_4)=(0.103,~0.178)$, d) $(E_4,E_5)=(0.178,~0.278)$, e) $(E_5,E_6)=(0.278,~0.403)$, and f) $(E_6,E_7)=(0.403,~\infty)$ (all in units of $\sigma_c^{-2}$), for width $a=\sqrt{2}\sigma_c$ of the initial wave packet and strength $V_0\sigma_c^{-2}=0.0325$ of the random potential. The numerical values have been extracted from the same data as the one used in Fig.~\ref{A} (red). Whereas good agreement is observed for low energies (a,b), differences are visible at larger energies (c-f), where the numerical data exhibits oscillations close to $x\simeq 0$ and $x\simeq\pm L/2$, which are not present in our theory. These oscillations can be traced back to interferences occurring due to the finite size of our system ($L=800\sigma_c$) and the periodic boundary conditions, as can be deduced by comparion with the corresponding density of a free particle in the same energy windows (blue line).}
\end{center}
\end{figure*}
\end{widetext}

To explain the reason for these small differences, we split the average asymptotic density into different energy intervals, i.e., we define:
\begin{equation}
\label{nintervalnum}
\overline{n^{(i)}(x)}=\overline{\sum_{n=1}^\infty |\langle x|\phi_n\rangle\langle\phi_n|\psi_0\rangle|^2\theta(E_n-E_i)\theta(E_{i+1}-E_n)},
\end{equation}
with Heaviside function $\theta(E)$. Eq.~(\ref{nintervalnum})  is similar to Eq.~(\ref{asymdefeigenfunctions}), except for the fact that the sum is restricted to eigenstates with energies $E_n\in [E_i,E_{i+1}]$ inside a certain energy interval. We choose
%tw
$E_1=-\infty$, $E_2=0$, $E_3=0.103$, $E_4=0.178$ and $E_5=0.278$, $E_6=0.403$ and $E_7=\infty$ (all in units of $\sigma_c^{-2}$), such that $\overline{n(x)}=\sum_{i=1}^6 \overline{n^{(i)}(x)}$.
The corresponding theoretical expression is constructed in a similar way from Eq.~(\ref{densitysimple}):
\begin{equation}
\overline{n^{(i)}(x)}=\int_{E_i}^{E_{i+1}} {\rm d}E~P(E)~\sum_{j=-\infty}^\infty \overline{n_{E}(x+j L)}.
\label{nintervaltheo}
\end{equation}
Fig.~\ref{ver1}(a-f) shows the corresponding comparison between theory and numerics in these four intervals. We observe very good agreement for negative energies, see Fig.~\ref{ver1}(a), where the asymptotic density is strongly localized due to the presence of bound states in the random potential. This result is remarkable for two reasons: first, it shows that the deviations between the theoretical  and the numerical energy distribution  $P(E)$, see Sec.~\ref{subsec:result_PE}, have no significant impact  on the spatial density profile. Second, remember that our theoretical approach is based on a diagrammatic method which takes into account only a certain class of diagrams (called \lq essential diagrams\rq\ in Sec.~\ref{subsec:berezinskii}). As explained in the original article of Berezinskii \cite{Berezinskii}, this restriction is, a priori, justified only for large energies. As our results show, however, the resulting expression for the spatial density profile $\overline{n_E(x)}$, Eq.~(\ref{eigenlocstate}), is valid also for low (and even negative) energies -- provided that an accurate value of the Lyapunov exponent $\gamma(E)$ is used in Eq.~(\ref{eigenlocstate}). 

A similar reasoning applies to the interval of small positive energies, see Fig.~\ref{ver1}(b). Deviations between theory and numerics become apparent, however, for larger energies, see  Fig.~\ref{ver1}(c-f). These can be traced back to the finite system size. 
Indeed, our simple way of taking into account the periodic boundary conditions, see the sum over $j$ in Eq.~(\ref{densitysimple}), after first having calculated the density according to a theory which is valid for an infinite system, neglects the occurrence of interferences in the finite, periodic system, e.g. interference between amplitudes of paths which propagate to the right-hand side ($x=L/2$) or to the left-hand side ($x=-L/2$) respectively, or between paths which perform several \lq loops\rq\ inside the periodic system. Such processes become relevant at high energies where the influence of the random potential becomes less important, and the particle behaves approximately like a free particle. To confirm this explanation, we also show the asymptotic density of a free particle within the corresponding energy intervals, where these interferences are clearly visible.

\section{Conclusions}
\label{sec:conclusion}

In this paper, we study the spatial expansion of an initially strongly confined wave packet in a one-dimensional correlated random potential. In particular, we present a theoretical description of the 
%tw average 
localized %tw
density 
%tw 
profile %tw
which is asymptotically reached at long times. In contrast to previous works \cite{Sanchez,Piraud}, our theory is able to explain, without adjustable parameters, not only the wings of the spatial density profile at large distances, but also its center in the vicinity of the initial state's position. Our theoretical description is based on a simple relation, Eq.~(\ref{density}), between the asymptotic density of the wave packet and the density autocorrelation function of energy eigenstates, Eq.~(\ref{asymdefeigenfunctionsberezinskii}), which takes into account the position and momentum uncertainty of the initial wave packet through its Wigner function, as well as its energy distribution through the spectral function $A(p,E)$. Whereas a similar relation has already been postulated in \cite{Piraud}, the present paper presents a 
diagrammatic derivation which puts its use on firm theoretical grounds. Our derivation uses the assumption  of an initially strongly confined wave packet. Therefore, its applicability to more extended initial states is still an open issue. We note that  the opposite extreme case of a wave packet initially strongly confined in momentum space has been treated in \cite{Kean}. As shown by numerical simulations, the asymptotic momentum distribution exhibits, in this case, two distinct peaks (a coherent backscattering and a coherent forward 
scattering peak \cite{Karpiuk}), and can also be described in terms of the autocorrelation function mentioned above.

As fundamental ingredients required in order to evaluate the asymptotic density profile, we develop analytical descriptions of the spectral function (using a variant of the self-consistent Born approximation), and of the Lyapunov exponent (using known  results \cite{Halperin,Thouless,Thouless2} which are exact for the case of white noise) for random potentials with short correlation length. More precisely, our analytical expression of the Lyapunov exponent is valid if the fluctuations of the random potential are smaller than the kinetic energy of a 
%tw momentum eigenstate exhibiting a
particle with %tw
wave length comparable to the correlation length ($V_0\sigma_c^2\ll 1$). Inserting this expression of the Lyapunov exponent into Gogolin's formula for the density autocorrelation function at fixed energy \cite{Gogolin2}, we observe remarkable agreement with numerical data  also at low (and even negative) energies, despite the fact that the underlying diagrammatic approach of Berezinskii \cite{Berezinskii} is justified only at large energies.

The theory developed in this paper will 
%tw
not only %tw
be useful for comparison with experiments on Anderson localization in one-dimensional disordered systems,
%tw Furthermore, 
but %tw
the improved theoretical insight based on the extension of Berezinskii's diagrammatic technique to the case of wave packets may 
%tw
also %tw
serve as a new starting point to explain the spreading of waves in nonlinear disordered systems \cite{Boris,Flach},
%tw
and thereby to clarify the impact of nonlinearity on the phenomenon of Anderson localization.
Given a precise, microscopic understanding of the underlying scattering processes in terms of diagrams, it appears to be promising to include nonlinearities %tw
in a similar way as it has been accomplished in order to describe 
%tw
the impact of nonlinearity \cite{Thomas1,Thomas2} or interactions between bosons \cite{Tobias1,Tobias2} on
coherent backscattering.\\
%tw in the presence of disorder and nonlinearity. 

\section{acknowledgments}
This work has been supported by Deutscher Akademischer Austausch Dienst (DAAD). The calculations have been performed on the Black Forest Grid Freiburg. We thank Alberto Rodriguez for a critical reading of the manuscript, and Andreas Buchleitner for useful discussions.

\appendix

\section{Derivation of Eq.~(\ref{density})} 
\label{sec:diagrammaticderivation}

We first present Berezinskii's and Gogolin's calculation of 
\begin{equation}
\Phi(x,x',E,\omega)=\overline{G^{(+)}(x-x',E+\omega)G^{(-)}(x'-x,E)}\label{eq:Phidef}
\end{equation}
 for a single source point $x'$, before we discuss its modification for the case of a wave packet with different source points $x''$ and $x'''$.

\subsection{Berezinskii's method}
\label{subsec:berezinskii_detailed}

To clarify the diagrammatic notation, we first give the explicit expression of the diagram shown in Fig.~\ref{fig:berezinskii}(a):
\begin{widetext}
\begin{eqnarray}
& & \Phi(x,x',E,\omega)_{\rm Fig.~\ref{fig:berezinskii}(a)}=\nonumber\\
& = & \int_{-\infty}^{x'}{\rm d}x_1 \int_{-\infty}^{x'} {\rm d}x_1'\int_{x_1}^{x'}{\rm d}x_2 \int_{x_1'}^{x'} {\rm d}x_2'  \int_{x'}^{x}{\rm d}x_3 \int_{x'}^{x} {\rm d}x_3' \int_{x_3}^{x}{\rm d}x_4 \int_{x_3'}^{x} {\rm d}x_4'\int_{x}^{\infty}{\rm d}x_5 \int_{x}^{\infty} {\rm d}x_5'~G_0^{(+)}(x_3-x',E+\omega)\nonumber\\
 & & \times G_0^{(+)}(x_2-x_3,E+\omega)G_0^{(+)}(x_1-x_2,E+\omega)G_0^{(+)}(x_2-x_1,E+\omega)G_0^{(+)}(x_4-x_2,E+\omega)G_0^{(+)}(x_3-x_4,E+\omega) \nonumber\\
 & & \times G_0^{(+)}(x_5-x_3,E+\omega)G_0^{(+)}(x-x_5,E+\omega)G_0^{(-)}(x_4'-x',E)G_0^{(-)}(x_1'-x_4',E)G_0^{(-)}(x_5'-x_1',E)G_0^{(-)}(x-x_5',E)
  \nonumber\\
 & &  \times C_2(x_1-x_1')C_2(x_2-x_2')C_2(x_3-x_3')C_2(x_4-x_4')C_2(x_5-x_5')
 \label{eq:example_berezinskii}
 \end{eqnarray}
 \end{widetext}
This term contributes to the average product $\Phi$ of Green functions defined in Eq.~(\ref{eq:Phidef}). 
As explained in Sec.~\ref{subsec:berezinskii}, it consists of free-particle Green functions $G_0^{(\pm)}$, see Eq.~(\ref{eq:G0}), and two-point correlation functions $C_2$, see Eq.~(\ref{correlationx}), of the random potential. Note that, for simplicity, the points $x_1',\dots,x_5'$ are not indicated in Fig.~\ref{fig:berezinskii}(a). For the case of short-range correlations (i.e. small correlation length $\sigma_c$), which we will assume in the following, these points are very close to the points $x_1,\dots,x_5$, with which they are correlated.

We now break up  each Green function into two factors associated with the respective  points connected by the Green function, e.g.
\begin{equation}
G_0^{(-)}(x_4'-x',E) = \left(\frac{i}{2 p_{E}}\right)^{1/2}e^{-ip_{E}x_4'}
 \times  \left(\frac{i}{2 p_{E}}\right)^{1/2}e^{ip_{E}x'}
\end{equation}
for $x_4'>x'$. We then perform the integrals over $x_1',\dots,x_5'$. Assuming a small correlation length $\sigma_c$, we may extend the limits of these integrations to $\pm\infty$. For small $\omega$, Eq.~(\ref{eq:example_berezinskii}) then turns into:
\begin{eqnarray}
& & \Phi(x,x',E,\omega)_{\rm Fig.~\ref{fig:berezinskii}(a)}\nonumber\\
& = & \frac{e^{-i\omega\frac{x+x'}{2p_E}}}{4 p_E^2}\int_{-\infty}^{x'}{\rm d}x_1\int_{x_1}^{x'}{\rm d}x_2 \int_{x'}^{x}{\rm d}x_3 \int_{x_3}^{x}{\rm d}x_4\int_{x}^{\infty}{\rm d}x_5
\nonumber\\
& & \times \frac{e^{-i\omega x_1/p_E}}{\ell_-} \left(-\frac{1}{\ell_+}\right) \left(-\frac{1}{\ell_-}\right) \frac{e^{i\omega x_4/p_E}}{\ell_-}\frac{e^{i\omega x_5/p_E}}{\ell_-}\label{eq:example_berezinskii2}
\end{eqnarray}
where the mean free paths $\ell_+$ (for forward scattering ) and  $\ell_-$ (for backward scattering) are introduced as follows:
\begin{subequations}
\begin{eqnarray}
\label{meanfreepath}
\frac{1}{\ell_+} & = & \frac{1}{2p_E^2}\int_0^{\infty}{\rm d}x\ C_2(x)\label{eq:ellplus}\\
\frac{1}{\ell_-}  \pm\frac{i}{\ell_0}& = & \frac{1}{2p_E^2}\int_0^{\infty}{\rm d}x\ C_2(x)e^{\pm 2ip_Ex}\label{eq:ellminus}
\end{eqnarray}
\end{subequations}
Note that $1/\ell_-=2 \gamma_{Born}(E)$ yields two times the Lyapunov exponent in Born approximation, see Eq.~(\ref{bornlyapunov}).

\begin{figure}
\includegraphics[width=8.5cm]{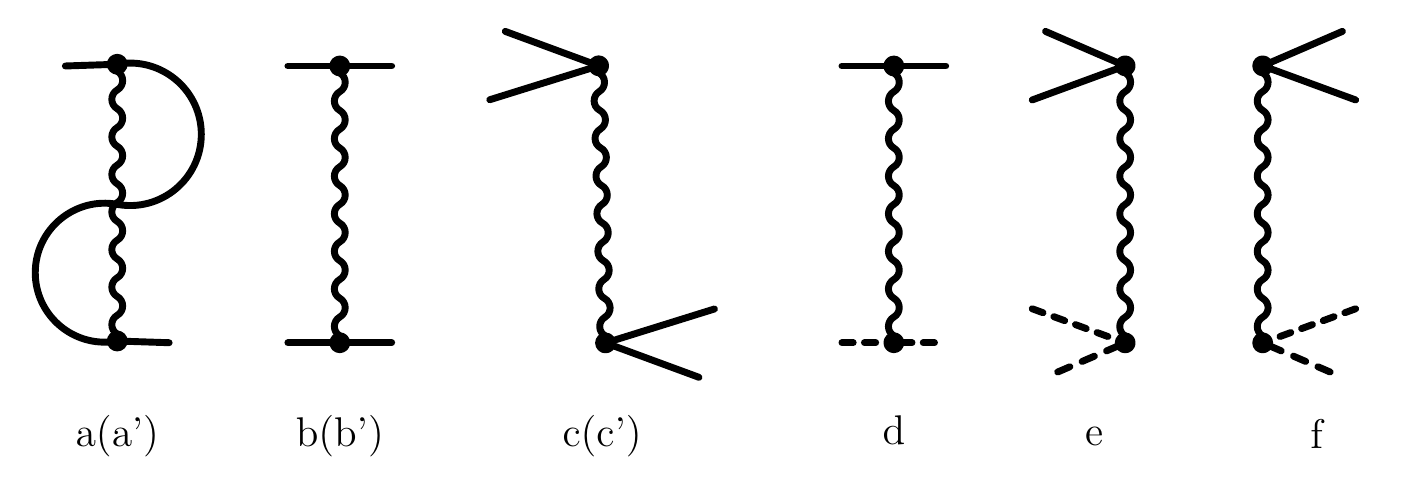}
\caption{Types of vertices entering in the essential diagrams \cite{Gogolin}. Solid (dashed) lines represent $G_0^{(\pm)}$, whereas two dots connected by a wavy line denote the two-point correlation function $C_2$. Diagrams (a',b',c') are the same as (a,b,c), but with solid lines replaced by dashed lines. The vertices correspond to the following factors: a) $-1/2\ell_--1/2\ell_+-i/2\ell_0$, a')  $-1/2\ell_--1/2\ell_++i/2\ell_0$ b,b') $-1/\ell_+$, c,c') $-1/\ell_-$, d) $1/\ell_+$, e) $e^{i\omega y/p_E}/\ell_-$, f) 
$e^{-i\omega y/p_E}/\ell_-$.
\label{fig:vertices}
}
\end{figure}

Each \lq essential diagram\rq\ (see Sec.~\ref{subsec:berezinskii}) contributing to the average product $\Phi$ 
contains certain types of vertices, which are displayed in Fig.~\ref{fig:vertices}. In the above example, a vertex of type f) is present at $x_1$, b) at $x_2$, c) at $x_3$ and e) at $x_4$ and $x_5$. In the second line of (\ref{eq:example_berezinskii2}), we recognize the terms accociated to each of these vertices. Additionally, (\ref{eq:example_berezinskii2}) also contains the factors associated to the initial and final points $x'$ and $x$. 

Let us now assume $x>x'$. In each essential diagram, we then distinguish the following three parts: the left-hand part lying to the left of $x'$, the right-hand part lying to the right of $x$, and the central part between $x'$ and $x$. Note that the left and right parts always contain an even number of solid (and of dashed) lines, and the central part an odd number.  We define $\tilde{R}_m(x)$ ($\tilde{L}_{m'}(x)$) as the sum of all right-hand (left-hand) diagrams containing $2m$ ($2m'$) solid and dashed lines at the boundary with the central part. Similarly, $Z_{m',m}(x',x)$ denotes the sum of all central parts with $2m+1$ ($2m'+1$) solid and dashed lines at the boundary with the right-hand (left-hand) part. The factors corresponding to the points $x$ and $x'$ are treated separately and not included in $\tilde{R}$, $Z$, or $\tilde{L}$. With this convention (which is slightly different from the one in Berezinskii's paper, where these factors are partly included in $Z$), the average product of Green functions results as follows: 

\begin{equation}
\Phi(x,x',E,\omega)=\Phi_R(x,x',E,\omega)+\Phi_L(x,x',E,\omega)
\label{eq:GGtotal}
\end{equation}
where

\begin{widetext}
\begin{eqnarray}
\Phi_R(x,x',E,\omega)
& = & 
\frac{e^{-\frac{i \omega x'}{2p_E}}}{4 p_E^2}
\sum_{m,m'=0}^{\infty} \tilde{L}_{m'}(x')Z_{m',m}(x',x)
\left(e^{\frac{i \omega x}{2p_E}}
\tilde{R}_m(x)+e^{-\frac{i \omega x}{2p_E}}
\tilde{R}_{m+1}(x)\right)
\label{eq:GGtotalR}
\\
\Phi_L(x,x',E,\omega)
& = & \frac{e^{\frac{i \omega x'}{2p_E}}}{4 p_E^2}
\sum_{m,m'=0}^{\infty} \tilde{L}_{m'+1}(x')Z_{m',m}(x',x)\left(e^{\frac{i \omega x}{2p_E}}
\tilde{R}_m(x)+e^{-\frac{i \omega x}{2p_E}}
\tilde{R}_{m+1}(x)
\right)
\label{eq:GGtotalL}
\end{eqnarray}
\end{widetext}
The two contributions, Eqs.~(\ref{eq:GGtotalR},\ref{eq:GGtotalL}), correspond to the cases where the 
particle leaves the initial point $x'$ towards the right-hand or left-hand side, respectively. Each of these contributions, in turn, is a sum of two terms 
corresponding to the particle arriving at the final point $x$ from the left-hand or right-hand side, respectively.
Fig.~\ref{fig:berezinskii}(a), for example, contributes to the second term $\tilde{L}_1(x')Z_{10}(x',x)\tilde{R}_1(x)$ in Eq.~(\ref{eq:GGtotalR}) (i.e. $m=1$ and $m'=0$). As shown in \cite{Berezinskii}, the two contributions, Eqs.~(\ref{eq:GGtotalR},\ref{eq:GGtotalL}), are identical in the limit $\omega\to 0$ 
%tw
(which is associated with the limit $m,m'\to\infty$, where the difference between $\tilde{L}_{m'}(x')$ and $\tilde{L}_{m'+1}(x')$ can be neglected). %tw 

Differential equations for $\tilde{R}$, $\tilde{L}$ and $Z$ can be obtained by considering infinitesimal shifts of the initial and final points $x'$ and $x$, and counting all possibilities of inserting one of the vertices displayed in Fig.~\ref{fig:vertices} in the corresponding infinitesimal interval. It is easy to verify that all terms associated with $1/\ell_+$ and $1/\ell_0$ counterbalance each other: let us denote the number of solid and dashed lines by $k$ (where $k=2m$ in the case of $\tilde{R}$ and $\tilde{L}$, whereas $k=2m+1$ in the case of $Z$). We then may attach the vertex (a) to any of the $k$ solid lines, and the vertex (a') to any of the $k$ dashed lines. Similarly, vertex (b) (or (b')) may be attached to 
$k(k-1)/2$ different pairs of two solid (or two dashed) lines, and vertex (d) to $k^2$ different pairs of one solid and one dashed line.
In total, this amounts to $k (-1/2\ell_--1/2\ell_+-i/2\ell_0-1/2\ell_--1/2\ell_++i/2\ell_0)- k(k-1)/\ell_+ + k^2/\ell_+=-k/\ell_-$, 
%tw 
i.e. all terms with $\ell_+$ and $\ell_0$ vanish. %tw
We are left with vertices (c), (c'), (e) and (f). Fig.~\ref{fig:insertion}(a) shows an insertion of vertex (f) in the interval $[x-{\rm d}x,x]$, which amounts to a change of the index $m$  from $\tilde{R}_{m+1}(x)$ to $\tilde{R}_m(x-{\rm d}x)$ (where $m=2$ in this example). As explained in Fig.~\ref{fig:insertion}(b), there are in total $m^2$ different ways of introducing the vertex (f). Repeating the same analysis for the other vertices and for $Z$ instead of $\tilde{R}$, we arrive at the following set of differential equations:

\begin{widetext}
\begin{figure*}
\includegraphics[width=8.5cm]{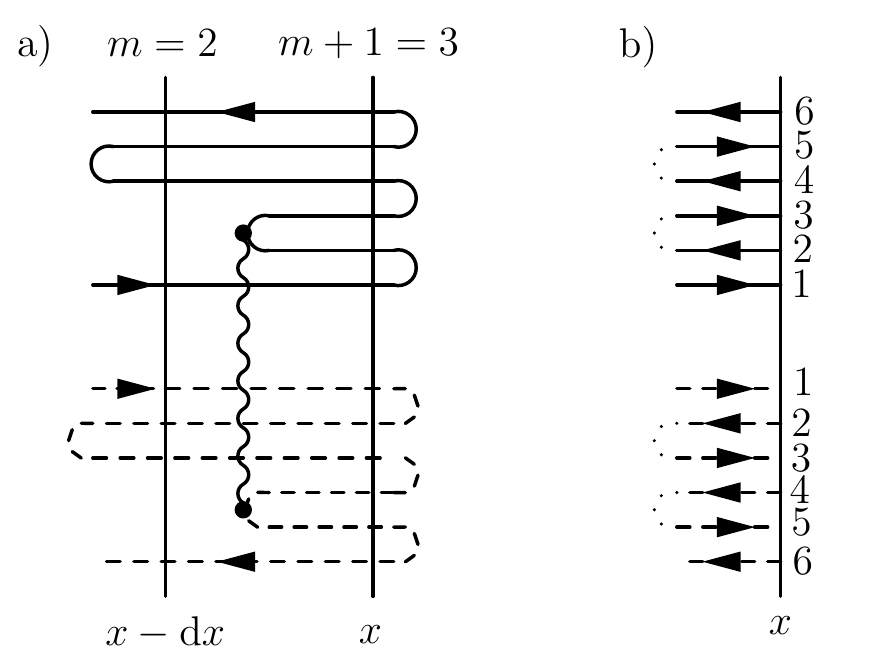}
\hspace*{0.5cm}
\includegraphics[width=7.5cm]{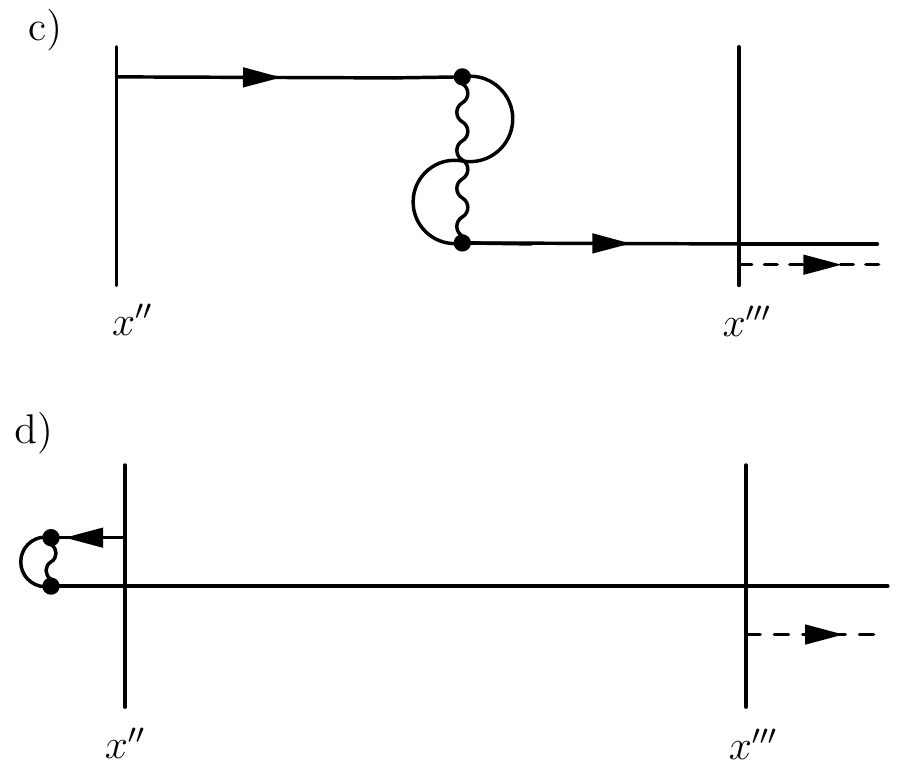}
\caption{a) The vertex displayed in Fig.~\ref{fig:vertices}(f) is inserted between $x$ and $x-{\rm d}x$ and thus changes the number of solid and dashed lines
from $2(m+1)=6$ at $x$ to $2m=4$ at $x-{\rm d}x$ (where $m=2$). All other vertices (outside this interval) are not displayed in the figure. 
b) Following the solid and the dashed line in the direction of the arrows indicated in a) (i.e. from the initial to the final point), we count the order in which the lines pass through the point $x$ from $1$ to $6$. The dotted lines indicate the places where vertex Fig.~\ref{fig:vertices}(f) can be inserted, such that the given order is respected. The example shown in a) results if the vertex is inserted between $2$ and $3$ for the solid lines and between $4$ and $5$ for the dashed lines, respectively. In total, there are $m^2=4$ different possibilities for inserting vertex 
Fig.~\ref{fig:vertices}(f). c) Inserting the vertex Fig.~\ref{fig:vertices}(a) on the line between $x''$ and $x'''$ yields the first-order expansion (in $x'''-x''$) of the term $\exp[i\tilde{p}_E(x'''-x'')]$.
d) The denominator $1/\tilde{p}_E$ present in the average Green function, see Eq.~(\ref{eq:Gposition}) is recovered by diagrams where the vertex occurs outside the interval $[x'',x''']$.
\label{fig:insertion}
}
\end{figure*}
\begin{subequations}
\begin{eqnarray}
-\frac{{\rm d}\tilde{R}_m}{{\rm d}x} & = & \frac{1}{\ell_-}\left(m^2 \tilde{R}_{m-1} e^{i \omega x/p_E}+m^2 \tilde{R}_{m+1} e^{-i \omega x/p_E}-2m^2 \tilde{R}_m\right)\label{eq:berezinskiiR}\\
\frac{{\rm d}Z_{m',m}}{{\rm d}x} & = & \frac{1}{\ell_-}\left(m^2 Z_{m',m-1} e^{-i \omega x/p_E}+(m+1)^2 Z_{m',m+1} e^{i \omega x/p_E}-\left(m^2+(m+1)^2\right) Z_{m',m}\right)\label{eq:berezinskiiZ}
\end{eqnarray}
\end{subequations}
\end{widetext}

The left-hand part follows through the symmetry relation $\tilde{L}_m(x)=\tilde{R}_m(-x)$. Furthermore, due to translational symmetry, $\tilde{R}_m(x)$ can be shown to fulfill $\tilde{R}_m(x)=e^{i\omega m x/p_E}R_m$ with position-independent coefficients $R_m$, which are given by
\begin{equation}
i\omega\ell_- R_m+m\left(R_{m+1}+R_{m-1}-2R_m\right)\label{eq:berezinskiiR}
\end{equation}
for $m\geq 1$ and $R_0=1$ (trivial multiplication with $1$ if no vertices are present in the right-hand part). Eq.~(\ref{eq:berezinskiiZ}) is supplemented with the boundary condition $Z_{m',m}(x',x)=\delta_{m',m}$ for $x=x'$. Eqs.~(\ref{eq:berezinskiiR}) and (\ref{eq:berezinskiiZ}) can now be solved by treating 
$m$ as a continous variable $p=-i m \omega\ell_-$ (which is justified in the limit $\omega\to 0$ where $m$ tends to infinity) and then solving differential equations in $p$ \cite{Berezinskii,Gogolin,Gogolin2}. The final result for the asymptotic density $\overline{n_E(x-x')}$, see Eq.~(\ref{eigenlocstate}), then follows from Eqs.~(\ref{eq:GGtotal}) and (\ref{bereft}). Furthermore, we note that the normalization factor, see Eq.~(\ref{bereft}), turns out as $\rho_E=1/(2\pi p_E)$. 

\subsection{Berezinskii's method for wave packets}
\label{subsec:berezinskii_wavepacket_detailed}

Let us now repeat the above steps in order to calculate 
\begin{equation}
\tilde{\Phi}(x,x'',x''',E,\omega)=\overline{G^{(+)}(x-x'',E+\omega)G^{(-)}(x'''-x,E)}
\end{equation}
 for the case of two different source points $x''$ and $x'''$. Let us first assume $x''<x'''<x$ and a particle initially propagating towards the right-hand side, as in the example shown in Fig.~\ref{fig:berezinskii}(b). Obviously, the difference with respect to the original Berezinskii diagram shown in Fig.~\ref{fig:berezinskii}(a) only concerns the
part between the source points $x''$ and $x'''$. Here, the number of solid lines ($2m+1$) and dashed lines ($2m$) is not identical. We thus introduce the quantity $A_m(x)$ describing this additional part. The differential equation for $A_m(x)$ can be obtained in a similar way as described above:
\begin{widetext}
\begin{equation}
\frac{{\rm d}A_m}{{\rm d}x}  =  \left[-\frac{1}{2\ell_-}-\frac{1}{2\ell_+}-\frac{i}{2\ell_0}-\left(m^2+(m-1)m\right)\frac{1}{\ell_-}\right]A_m+
\frac{1}{\ell_-}\left(m^2 A_{m-1} e^{-i \omega x/p_E}+m (m+1) A_{m+1} e^{i \omega x/p_E}\right)
\label{eq:Adifferential}
\end{equation}
As boundary condition, we impose $A_m(x'')=\tilde{L}_m(x'')$, since the additional part $A$ must be connected to the left-hand part $\tilde{L}$ at $x=x''$. Similarly to Eq.~(\ref{eq:GGtotal}), we obtain:
\begin{equation}
\tilde{\Phi}(x,x'',x''',E,\omega)=\tilde{\Phi}_R(x,x'',x''',E,\omega)+\tilde{\Phi}_L(x,x'',x''',E,\omega)
\label{eq:GGtotalwavepacket}
\end{equation}
where
\begin{equation}
\tilde{\Phi}_R(x,x'',x''',E,\omega)  = e^{ip_E(x'''-x'')}\frac{e^{-\frac{i\omega x''}{2p_E}}}{4 p_E^2}
  \sum_{m,m'=0}^{\infty} A_{m'}(x''')Z_{m',m}(x''',x)
\left(
e^{\frac{i\omega x}{2p_E}}
\tilde{R}_m(x)+ e^{-\frac{i \omega x}{2p_E}}
\tilde{R}_{m+1}(x)\right)
\label{eq:GGtotalwavepacketR}
\end{equation}
and  similarly  for $\tilde{\Phi}_L(x,x'',x''',E,\omega)$ (see below).
Let us now compare Eqs.~(\ref{eq:GGtotalR}) and (\ref{eq:GGtotalwavepacketR}). The essential difference consists of the term  $\tilde{L}_{m'}(x')Z_{m',m}(x',x)$ occurring in Eq.~(\ref{eq:GGtotalR}) instead of $A_{m'}(x''')Z_{m',m}(x''',x)$ in Eq.~(\ref{eq:GGtotalwavepacketR}). Expanding the solution of Eq.~(\ref{eq:Adifferential}) in first order of $x'''-x''$, we obtain:
\begin{eqnarray}
A_{m'}(x''') & = & \tilde{L}_{m'}(x'')+(x'''-x'') \Biggl\{\left[-\frac{1}{2\ell_-}-\frac{1}{2\ell_+}-\frac{i}{2\ell_0}-\left((m')^2+(m'-1)m'\right)\frac{1}{\ell_-}\right]\tilde{L}_{m'}(x'')+\Biggr.\nonumber\\
& & +\Biggl.
\frac{1}{\ell_-}\left((m')^2 \tilde{L}_{m'-1}(x'') e^{-i \omega x''/p_E}+m' (m'+1) \tilde{L}_{m'+1}(x'') e^{i \omega x''/p_E}\right)\Biggr\}
\end{eqnarray}
Similarly, we have:
\begin{eqnarray}
\tilde{L}_{m'}(x'') & = & \tilde{L}_{m'}(x')-\frac{x'-x''}{\ell_-}\left((m')^2 \tilde{L}_{m'-1}(x') e^{-i \omega x'/p_E}+(m')^2 \tilde{L}_{m'+1}(x') e^{i \omega x'/p_E}-2(m')^2 \tilde{L}_{m'}(x')\right)\label{eq:berezinskiiLdiff}\\
Z_{m',m}(x''',x) & = & Z_{m',m}(x',x)-\frac{x'''-x'}{\ell_-} \Bigl((m')^2 Z_{m'-1,m}(x',x) e^{i \omega x'/p_E}+(m'+1)^2 Z_{m'+1,m}(x',x) e^{-i \omega x'/p_E}\Bigr.\nonumber\\
& & \Bigl.-\left((m')^2+(m'+1)^2\right) Z_{m',m}(x',x)\Bigr)\label{eq:berezinskiiZdiff}
\end{eqnarray}
where Eq.~(\ref{eq:berezinskiiLdiff}) results from Eq.~(\ref{eq:berezinskiiR}) together with $\tilde{L}_m(x)=\tilde{R}_m(-x)$, and Eq.~(\ref{eq:berezinskiiZdiff}) from Eq.~(\ref{eq:berezinskiiZ}) together with $Z_{m',m}(x,x')=Z_{m,m'}(-x',-x)$. Summing over $m'$ and keeping again only terms linear in the difference $x'''-x''=2 (x'-x'')=2 (x'''-x')$ (since $x'=\frac{x''+x'''}{2}$), we obtain:
\begin{eqnarray}
\sum_{m'} A_{m'}(x''')Z_{m',m}(x''',x) & = & \sum_{m'}\tilde{L}_{m'}(x')Z_{m',m}(x',x)+
(x'''-x'')\left(-\frac{1}{2\ell_-}-\frac{1}{2\ell_+}-\frac{i}{2\ell_0}\right) \sum_{m'}\tilde{L}_{m'}(x')Z_{m',m}(x',x)\nonumber\\
& & +\sum_{m'} (x'''-x'') e^{-i\omega m x'/p_E} \frac{R_{m'+1}-R_{m'}}{2} Z_{m',m}(x',x)\label{eq:berezinskiitwoterms}
\end{eqnarray}
\end{widetext}

The term $R_{m'+1}-R_{m'}$ can be neglected in the limit $m'\to\infty$ corresponding to $\omega\to 0$ \cite{Berezinskii}.
Defining the effective complex wavevector
\begin{equation}
\tilde{p}_E=p_E +\frac{i}{2\ell_-}+\frac{i}{2\ell_+}-\frac{1}{2\ell_0}
\label{eq:ktilde}
\end{equation}
describing average propagation in the random potential, it follows from Eqs.~(\ref{eq:GGtotalR},\ref{eq:GGtotalwavepacketR},\ref{eq:berezinskiitwoterms}) that: 
\begin{equation}
\tilde{\Phi}_R(x,x'',x''',E,\omega) = e^{i\tilde{p}_E(x'''-x'')} \Phi_R(x,x',E,\omega)
\label{eq:Phirsolution}
\end{equation}
for small $\omega$ and small $x'''-x''$. The corresponding relation for $\tilde{\Phi}_L(x,x'',x''',E,\omega)$ can immediately be deduced from
Eq.~(\ref{eq:Phirsolution}), since an initially left-propagating diagram can be mapped to a right-propagating one by exchanging solid with dashed lines and inverting the coordinates, i.e. $\tilde{\Phi}_L(x,x'',x''',E,\omega) =\tilde{\Phi}_R^*(-x,-x''',-x'',E+\omega,-\omega)$ and
$\Phi_L(x,x',E,\omega) =\Phi_R^*(-x,-x',E+\omega,-\omega)$. From this, we obtain:
\begin{equation}
\tilde{\Phi}_L(x,x'',x''',E,\omega) = e^{-i\tilde{p}_E^*(x'''-x'')} \Phi_L(x,x',E,\omega)
\label{eq:Philsolution}
\end{equation}
Taking into account that, as mentioned above, $\Phi_R(x,x',E,\omega)=\Phi_L(x,x',E,\omega)$ (for small $\omega$), the sum of Eqs.~(\ref{eq:Phirsolution}) and (\ref{eq:Philsolution}) yields:
\begin{widetext}
\begin{equation}
\rho_E \overline{G^{(+)}(x-x'',E+\omega)G^{(+)}_E(x'''-x,E)} = \frac{e^{i\tilde{p}_E|x'''-x''|}+e^{-i\tilde{p}_E^*|x'''-x''|}}{4\pi p_E} 
\overline{G^{(+)}(x-x',E+\omega)G^{(+)}_E(x'-x,E)}\label{eq:berezinskiiwpfinal}
\end{equation}
\end{widetext}
where we included the normalization factor $\rho_E=1/(2\pi p_E)$ on both sides of the equation.
 This expression is valid also for $x'''<x''$ (as can be proven by inverting the sign of all spatial arguments and exchanging $L\leftrightarrow R$). 
 
 As explained in Sec.~\ref{subsec:berezinskiiwp}, our aim is to show that the prefactor
on the right-hand side of Eq.~(\ref{eq:berezinskiiwpfinal}) reproduces the imaginary part of the average Green function. 
According to the Dyson equation (see Sec.~\ref{subsec:spectraldensity}) the latter is related as follows
\begin{eqnarray}
\overline{G^{(+)}(x''-x''',E)} & = & G^{(+)}_0(x''-x''',E)\label{eq:dyson_position}\\
&  & \hspace{-3cm} + \int_{-\infty}^\infty {\rm d}\tilde{x}~G^{(+)}_0(x''-\tilde{x},E) \Sigma(E) \overline{G^{(+)}(\tilde{x}-x''',E)}\nonumber
\end{eqnarray}
to the self-energy $\Sigma(E)$, which, using the Born approximation, see Eq.~(\ref{eq:Sigmaborn1}), is given by $\Sigma(E)=p_E^2-\tilde{p}_E^2\simeq -2 p_E (\tilde{p}_E-p_E)$, with
$\tilde{p}_E$ as defined in Eq.~(\ref{eq:ktilde}). Diagrammatically, this self energy corresponds to vertex (a) displayed in Fig.~\ref{fig:vertices}. The solution of 
Eq.~(\ref{eq:dyson_position}) is given by Eq.~(\ref{eq:Gposition}),
which has the same form as the free-particle Green function, see Eq.~(\ref{eq:G0}), but with the wave number $p_E$ replaced by the effective wave number $\tilde{p}_E$. As compared to the prefactor in Eq.~(\ref{eq:berezinskiiwpfinal}), however, the denominator
$1/p_E$ is replaced by $1/\tilde{p}_E$. Although this difference is small in the validity regime of the diagrammatic approach ($p_E \ell_-\gg 1$), 
we will show in the following how the factor $1/\tilde{p}_E$ can be recovered by considering additional diagrams.

In our above derivation, we have restricted ourselves to diagrams where vertex (a) is inserted only in the interval between  $x''$ and $x'''$, as indicated in 
%tw 
Fig.~\ref{fig:insertion}(c). %tw
Here, we focus on the initial stage of propagation, i.e., we consider the solid line originating from the source point $x''$ until it passes the second source point $x'''$. Up to first order in $x'''-x''$, the contribution from these diagrams to the average Green function reads:
\begin{eqnarray}
 \overline{G^{(+)}(x''-x''',E)} & = & G^{(+)}_0(x''-x''',E)\nonumber\\
&  & \hspace{-3cm} + \int_{x''}^{x'''} {\rm d}\tilde{x}~G^{(+)}_0(x''-\tilde{x},E) \Sigma(E) G_0^{(+)}(\tilde{x}-x''',E)=
\nonumber\\
& & \hspace{-3cm}   =G_0^{(+)}(x''-x''',E)\left(1+i(\tilde{p}_E-p_E)(x'''-x'')\right) \nonumber\\
& & \hspace{-3cm} \simeq -\frac{i}{p_E} e^{i\tilde{p}_E (x'''-x'')}
\end{eqnarray}
which exhibits $p_E$ instead of $\tilde{p}_E$ in the denominator, as in Eq.~(\ref{eq:berezinskiiwpfinal}). 
According to Eq.~(\ref{eq:dyson_position}) -- where the integration over $\tilde{x}$ is not restricted to $[x'',x''']$ --
and its solution, Eq.~(\ref{eq:Gposition}),
the denominator
$1/\tilde{p}_E$ is recovered if the vertex (a) is inserted outside the interval $[x'',x''']$, as in the diagram depicted in 
%tw 
Fig.~\ref{fig:insertion}(d).

This yields our final result:
\begin{eqnarray}
\overline{n(x)} & = & \int_{-\infty}^\infty {\rm d}E \int_{-\infty}^\infty {\rm d}x''{\rm d}x'''  \overline{n_E\left(x-\frac{x''+x'''}{2}\right)}
\nonumber\\
&  & \hspace{-1cm} \times \langle x''|\psi_0\rangle\langle\psi_0|x'''\rangle\left(-\Im\left\{\overline{G^{(+)}(x''-x''',E)}\right\}\right)
\label{eq:berwp2}
\end{eqnarray}
where, as compared to Eq.~(\ref{eq:berwp1}), the free-particle Green function is replaced by the average Green function. This result is valid under the assumption of a strongly confined initial wave packet ($\sigma\ll \ell_-$), where the first-order expansion in the difference $x'''-x''$ between the two source points, which we applied several times in our above derivation, is justified.

\bibliography{gpl_Juan_Thomas.bib}
\end{document}